\documentclass[journal]{IEEEtran}
\pdfoutput=1 
\usepackage{amsfonts}
\usepackage[cmex10]{amsmath}
\usepackage{amsfonts}
\usepackage[fleqn,tbtags]{mathtools}
\usepackage{cite}
\usepackage{mathtools}

\usepackage{multirow}
\usepackage[]{algorithm2e}
\usepackage{algorithmicx}
\usepackage{color}
\usepackage{verbatim}
\usepackage{subcaption,graphicx}

\def\tr{\mbox{${\mbox{tr}}$}}
\def\dashfill{\cleaders\hbox{-~-}\hfill}
\makeatletter
\newcommand*{\rom}[1]{\expandafter\@slowromancap\romannumeral #1@}
\makeatother

\ifCLASSINFOpdf
\else
\fi

\begin{document}
\title{On Distributed Vector Estimation for Power and Bandwidth Constrained Wireless Sensor Networks}

\author{Alireza~Sani,~\IEEEmembership{Student Member,~IEEE,}
        Azadeh~Vosoughi,~\IEEEmembership{Senior Member,~IEEE,}
\thanks{Part of this research is presented at 48th Asilomar Conference on Signals, Systems and Computers, 2014, and 34th Military Communication, 2015\cite{sani-asilomar2014,sani-milcom2015}. This research is supported by NSF under grants CCF-1336123, CCF-1341966, and CCF-1319770.
}
}


%
%

\markboth{Submitted to Journal of Transactions on Signal Processing,}%
{Shell \MakeLowercase{\textit{et al.}}: Bare Demo of IEEEtran.cls for Journals}
%



\maketitle


\begin{abstract} 
We consider distributed estimation of a Gaussian vector with a linear observation model in an inhomogeneous wireless sensor network, where a fusion center (FC) reconstructs the unknown vector, using a linear estimator.  Sensors employ uniform multi-bit quantizers and binary PSK modulation,  and communicate with the FC over orthogonal power- and bandwidth-constrained wireless channels.
We study transmit power and quantization rate (measured in bits per sensor) allocation schemes that minimize mean-square error (MSE). In particular, we derive two closed-form upper bounds on the MSE, in terms of the optimization parameters  and propose ``coupled'' and ``decoupled'' resource allocation schemes that minimize these bounds. We show that the bounds are good approximations of the simulated MSE and the performance of the proposed schemes approaches the clairvoyant centralized estimation when total transmit power or bandwidth is very large. We study how the power and rate allocation are dependent on sensors' observation qualities and channel gains, as well as total transmit power and bandwidth constraints. Our simulations corroborate our analytical results and illustrate the superior performance of the proposed algorithms.
%
%
\end{abstract}


%
\IEEEpeerreviewmaketitle

\vspace{-0.4cm}
\section{Introduction}
Distributed parameter estimation problem for wireless sensor networks (WSNs) has a rich literature in signal processing community. Several researchers study quantization design, assuming that sensors' observations are sent over bandwidth constrained (otherwise error-free) communication channels, examples are
%
%
\cite{varshney2014-quantizer, Giannakis2006.1.tsp, Giannakis2006tsp, willett-tsp-2006, AlRegib2007tsp, Varshney-2010, varshney-tsp-2012}.
In particular, \cite{varshney2014-quantizer} designs the optimal quantizers which maximize Bayesian Fisher information, for estimating a random parameter.
Assuming identical one-bit quantizers, \cite{Varshney-2010, varshney-tsp-2012} find the minimum achievable Cram\'{e}r-Rao lower bound (CRLB) and the optimal quantizers, for estimating a deterministic parameter.
%
%
The authors in \cite{Giannakis2006.1.tsp, Giannakis2006tsp} investigate one-bit quantizers for estimating a deterministic parameter, when the FC employs maximum-likelihood estimator. 
For estimating a random parameter with uniform variable rate quantizers, 
\cite{willett-tsp-2006} studies the tradeoff between quantizing a few sensors finely or as many sensors as possible coarsely and its effect on Fisher information, subject to a total rate constraint. 
%
%
For estimating a deterministic parameter \cite{AlRegib2007tsp} investigates a bit allocation scheme that minimizes the mean square error (MSE) when the FC utilizes best linear unbiased estimator (BLUE), subject to a total bit constraint.  
%
%
Several researchers relax the assumption on communication channels being error-free \cite{valenti.tsp.2013, Nevat-Collings-2013}.
%
%
For estimating a vector of deterministic parameters, \cite{valenti.tsp.2013} studies an expectation-maximization algorithm and the CRLB, when sensors employ fixed and identical multi-bit quantizers and communication channel model is additive white Gaussian noise (AWGN). 
%
%
%
A related problem is studied in \cite{Nevat-Collings-2013}, in which the FC employs a spatial BLUE for field reconstruction and the MSE is compared with a posterior CRLB.

Recent years have also witnessed a growing interest in studying energy efficient distributed parameter estimation for WSNs, examples are \cite{goldsmith2006, AlRegib2009tsp, Giannakis-2008}. 
%
%
%
For estimating a deterministic parameter \cite{goldsmith2006} explores the optimal power allocation scheme which minimizes total transmit power of the network, subject to an MSE constraint, when  the FC utilizes BLUE and sensors digitally transmit to the FC over orthogonal AWGN channels.
%
%
A converse problem is considered in \cite{AlRegib2009tsp},  where the authors minimize the MSE, subject to a total transmit power constraint. For a homogeneous WSN, \cite{Giannakis-2008} investigates a bit and power allocation scheme that minimizes the MSE, subject to a total transmit power constraint, when communication channels are modeled as binary symmetric channels. We note that \cite{goldsmith2006, AlRegib2009tsp, Giannakis-2008} do not include a total bit constraint in their problem formulations. 
%
There is a collection of elegant results on energy efficient distributed parameter estimation with analog transmission, also known as  Amplify-and-Forward (AF), examples are \cite{Khandani-2008, Jun-Fang-2009, jafarkhani-tsp-2014, Jafarkhani2012tsp, Turkiyyah-tsp-2013}. However, the problem formulations in these works naturally cannot have a bandwidth constraint.
%

Different from the aforementioned works, that consider either transmit power or bandwidth constraint, we consider distributed parameter estimation, subject to both total transmit power and bandwidth constraints. 
Similar to \cite{Giannakis2006.1.tsp, Giannakis2006tsp, AlRegib2007tsp}, we choose total number of bits allowed to be transmitted by all sensors to the FC as the measure of total bandwidth.
From practical perspectives, having a total transmit power constraint 
enhances energy efficiency in battery-powered WSNs.
Putting a cap on total bandwidth can further improve energy efficiency, since data communication is a major contributor to the network energy consumption.
Our new constrained problem formulation allows us to probe the impacts of both constraints on resource allocation and overall estimation accuracy, and enables us to find the best resource allocation (i.e., power and bit) in extreme cases where: (i) we have scarce total transmit power and ample total bandwidth, (ii) we have plentiful total transmit power and scarce total bandwidth.
%
The paper organization follows. Section \ref{our-system-model} introduces our system model and set up our optimization problem. In Section \ref{section-characterization-MSE} we derive two closed-form upper bounds, $\mathcal{D}_a$ and $\mathcal{D}_b$, on the MSE corresponding to  the linear estimator at the FC, in terms of the optimization parameters (i.e., transmit power and quantization rate per sensor). In Section \ref{optimization-algorithms} we propose ``coupled'' resource allocation schemes that minimize these bounds, utilizing the iterative ellipsoid method. This method conducts multi-dimensional search to find the quantization rate vector. In Section \ref{optimization-algorithms-decoupled} we propose ``decoupled'' resource allocation schemes, which rely on one-dimensional search to find the quantization rates. Section \ref{smulation-section} discusses our numerical results. Section \ref{conclusions} concludes our work. 

%
\begin{figure}[t]
\centering
\includegraphics[width=3.5in]{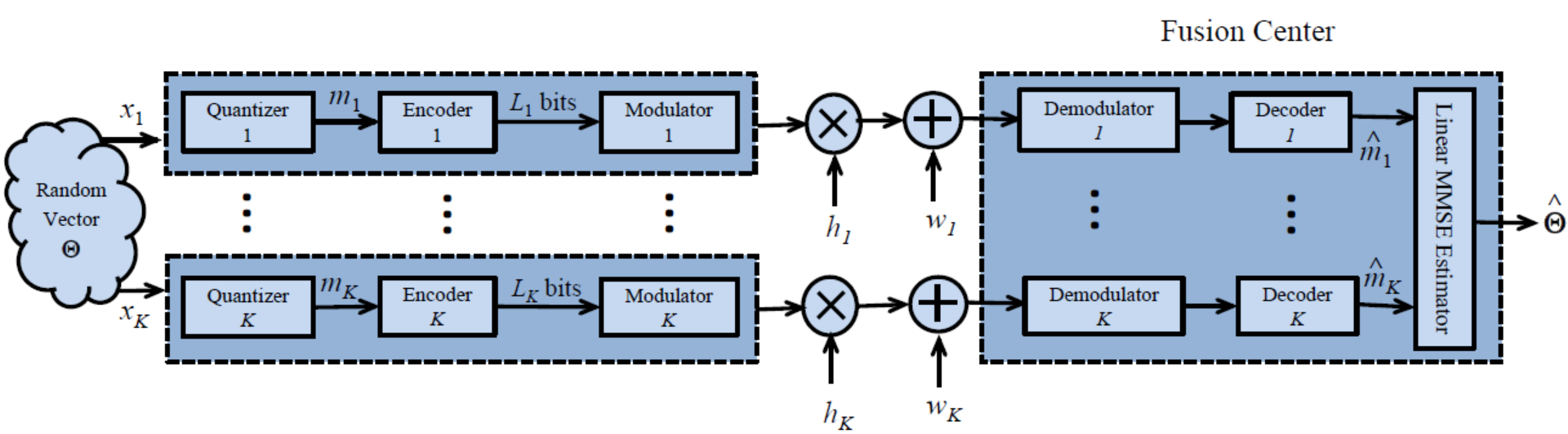}
\caption{Our system model consists of $K$ sensors and a FC, tasked with estimating a Gaussian vector  $\boldsymbol \theta$.}
\label{fig0}
\vspace{-0.5cm}
\end{figure}
%
\section{System Model and Problem Statement}\label{our-system-model}

We consider a wireless network with $K$ spatially distributed inhomogeneous sensors. Each sensor makes a noisy observation, which depends on an unobservable vector of random parameters $\boldsymbol \theta$, processes locally its observation, and transmits a summary of its observation to a fusion center (FC) over erroneous wireless channels. The FC is tasked with estimating $\boldsymbol \theta$, via fusing the collective received data from sensors (see Fig.\ref{fig0}).
We assume 
$\boldsymbol \theta\!=\![\theta_1, ...,\theta_q]^T \! \in \!\mathbb{R}^q$ is zero-mean Gaussian with covariance matrix
$\mathcal{C}_{\boldmath{\theta}}\!=\! \mathbb{E}\{\boldsymbol{\theta\theta^T}\}$.
Let random scalar $x_k$ denote the noisy observation of sensor $k$. We assume the following linear observation model:
\begin{equation} \label{observation model}
x_k=\mathbf{a}_k^T \boldsymbol{\theta}  +n_k, ~~~  k=1,...,K,
\end{equation}
where $\textbf{a}_k\!=\![{a_k}_1, ..., {a_k}_q]^T \! \in \! \mathbb{R}^q$ is known observation gain vector and $n_k \! \sim \! {\cal N}(0,\sigma_{n_k}^2)$ is observation noise. We assume $n_k$'s are mutually uncorrelated and also are uncorrelated with {\boldmath$\theta$}. Define observation vector 
$\boldsymbol{x}\!=\![{x}_1,...,{x}_K]^T$, matrix $\mathbf{A}\!=\![\mathbf{a}_1,...,\mathbf{a}_K]$, and diagonal covariance matrix $\mathcal{C}_{\boldsymbol{n}} \!=\!\mbox{diag}(\sigma_{n_1}^2,...,\sigma_{n_K}^2)$. Suppose $\mathcal{C}_{{\boldsymbol{x}}}\!=\!\mathbb{E}\{{\boldsymbol{x}}   {\boldsymbol{x}}^T \}$ and $\mathcal{C}_{{\boldsymbol{x}\boldsymbol{\theta}}}\!=\!\mathbb{E}\{{\boldsymbol{x}}   {\boldsymbol{\theta}}^T \}$, respectively, represent the covariance matrix of $\boldsymbol{x}$ and cross-covariance matrix of $\boldsymbol{x}$ and $\boldsymbol{\theta}$. It is easy to verify:
\begin{equation} \label{covariance matrices}
\mathcal{C}_{{\boldsymbol{x}}{\boldsymbol{\theta}}} =\mathbf{A}^T\mathcal{C}_{\boldsymbol{\theta}}, ~~~ \mathcal{C}_{{\boldsymbol{x}}} =\mathbf{A}^T\mathcal{C}_{\boldsymbol{\theta}}\mathbf{A}+\mathcal{C}_{\boldsymbol{n}}.\nonumber
\end{equation}
Sensor $k$ employs a uniform quantizer with $M_k$ quantization levels and quantization step size $\Delta_k$, to map $x_k$ into a quantization level $m_k \! \in \! \{m_{k,1},..., m_{k,M_k} \}$, where $m_{k,i}\!=\!\frac{(2i-1-M_k)\Delta_k}{2}$ for $i\!=\!1, ..., M_k$. 
Considering our observation model, we assume $x_k$ lies in the interval $[-\tau_k, \tau_k]$ almost surely, for some reasonably large value of $\tau_k$, i.e., the probability $p(|x_k| \! \geq \! \tau_k) \! \approx \! 0$. Consequently, we let  $\Delta_k\!=\!\frac{2\tau_k}{M_k -1}$. These imply that, the quantizer maps $x_k$ as the following: if $x_k  \! \in \! [m_{k,i}-\frac{\Delta_k}{2}, m_{k,i}+\frac{\Delta_k}{2}]$, then $m_k\!=\!m_{k,i}$, if $x_k \geq \tau_k$, then $m_k \! =\! \tau_k$, and if $x_k  \! \leq \! -\tau_k$, then $m_k \!=\!-\tau_k$. Following quantization, sensor $k$ maps the index $i$ of $m_{k,i}$ into a bit sequence of length $L_k\!=\!\log_2 M_k$, and finally modulates these $L_k$ bits into $L_k$ binary PSK (BPSK) modulated symbols\cite{Giannakis-2008}. 
Sensors send their symbols to the FC over orthogonal wireless channels, where transmission is subject to both transmit power and bandwidth constraints. The $L_k$ symbols sent by sensor $k$ experience flat fading with a fading coefficient $h_k$ and are corrupted by a zero mean complex Gaussian receiver noise $w_k$ with variance $\sigma_{w_k}^2$. 
We assume $w_k$'s are mutually uncorrelated and $h_k$  does not change during the transmission of $L_k$ symbols. Let $P_k$ denote the transmit power corresponding to $L_k$ symbols from sensor $k$, which we assume it is distributed equally among $L_k$ symbols. Suppose there are constraints on the total transmit power and bandwidth of this network, i.e., $\sum_{k=1}^K P_k \! \leq \! P_{tot}$ and $\sum_{k=1}^K L_k \! \leq \! B_{tot}$. 

In the absence of knowledge of the joint distribution of $\boldsymbol{x}$ and $\boldsymbol{\theta}$, we resort to linear estimators \cite{jafarkhani-tsp-2014, AlRegib2009tsp}, that only require knowledge of $\mathbf{A}$ and second-order statistics $\mathcal{C}_{{\boldsymbol{\theta}}}$ and $\mathcal{C}_{{\boldsymbol{n}}}$, to estimate $\boldsymbol{\theta}$ with low computational complexity.
To describe the operation of linear estimator at the FC, let  $\hat{m}_k$ denote the recovered quantization level corresponding to sensor $k$, where in general $\hat{m}_k \! \neq \! m_k$, due to communication channel errors. 
The FC first processes the received signals from the sensors individually, to recover the transmitted quantization levels. Having $\hat{m}_1,...,\hat{m}_K$, the FC applies a linear estimator to form the estimate $\hat{\boldsymbol{\theta}}$. Let ${\boldsymbol{\mathcal{D}}}_0\! =\! \mathbb{E}
\{(\hat{{\boldsymbol{\theta}}}-\boldsymbol{\theta})(\hat{{\boldsymbol{\theta}}}-\boldsymbol{\theta})^T\} $ denote
the error correlation matrix corresponding to this linear estimator, whose \textit{i}-th diagonal entry, $[{\boldsymbol{\mathcal{D}}}_0]_{\textit{i,i}}$,  is the MSE corresponding to the \textit{i}-th entry of vector $\boldsymbol{\theta}$. We choose $\mathcal{D}_0\! =\! \tr({\boldsymbol{\mathcal{D}}}_0)$ as our MSE distortion metric \cite{StevenKay1993}. 
Our goal is to find the optimal resource allocation scheme, i.e., quantization rate $L_k$ and transmit power $P_k$ $\forall k$, that minimize $\mathcal{D}_0$. In other words, we are interested to solve the following optimization problem: 
\begin{eqnarray} \label{original Problem}
 &&\underset{L_k,P_k, ~\forall k} {\text{minimize}} ~
 \mathcal{D}_0(\{L_k,P_k\}_{k=1}^K) \\
&& \text{s.t.}
\sum_{k=1}^K L_k\leq B_{tot}, \sum_{k=1}^K P_k\leq P_{tot}, L_k \in \mathbb{Z}_{+}, P_k \in \mathbb{R}_{+}, ~\forall k. \nonumber
\end{eqnarray}
\vspace{-0.6cm}
%
\section{Characterization of MSE}\label{section-characterization-MSE}

We wish to characterize $\mathcal{D}_0$ in terms of the optimization parameters $\{L_k,P_k\}_{k=1}^K$.
To accomplish this, we take a two-step approach \cite{goldsmith2006}: in the first step, we assume that the quantization levels  transmitted by the sensors are received error-free at the FC. Based on the error-free transmission  assumption, we characterize the
MSE, due to observation noises and quantization errors. In the second step, we take into account the contribution of wireless communication channel errors on the MSE. This approach provide us with an upper bound on $\mathcal{D}_0$, which can be expressed in terms of  $\{L_k,P_k\}_{k=1}^K$.
Define vector $\boldsymbol{m}\!=\![{m}_1,...,{m}_K]^T$, including transmitted quantization levels for all sensors, and vector $\hat{\boldsymbol{m}}\!=\![\hat{{m}}_1,...,\hat{{m}}_K]^T$, consisting of recovered quantization levels at the FC. Let: 
\begin{equation}\label{LMMSE-G-matrix}
\check{\boldsymbol{\theta}}\!=\!\boldsymbol{G}\boldsymbol{m},~~~\mbox{where}~~ \boldsymbol{G}\!=\!\mathbb{E}\{\boldsymbol{\theta}\boldsymbol{m}^T\}(\mathbb{E}\{\boldsymbol{m}\boldsymbol{m}^T\})^{-1} .
\end{equation}
Note that $\check{\boldsymbol{\theta}}$ is the linear minimum MSE (LMMSE) estimator, had the FC received the transmitted quantization levels error-free \cite{StevenKay1993}. Having $\hat{\boldsymbol{m}}$, the FC employs the same linear operator $\boldsymbol{G}$ to obtain the linear estimate $\hat{\boldsymbol{\theta}}\!=\!\boldsymbol{G}\hat{\boldsymbol{m}}$. To characterize the MSE, we define covariance matrices 
${\boldsymbol{\mathcal{D}}_1}\!=\!\mathbb{E}
\{(\check{{\boldsymbol{\theta}}}-\boldsymbol{\theta})(\check{{\boldsymbol{\theta}}}-\boldsymbol{\theta})^T\} $ and ${\boldsymbol{\mathcal{D}}_2}\!=\!\mathbb{E}
\{(\hat{{\boldsymbol{\theta}}}-\check{\boldsymbol{\theta}})(\hat{{\boldsymbol{\theta}}}-\check{\boldsymbol{\theta}})^T\} $. One can verify:
\begin{equation}\label{Distortion-UP-Matrix}
{\boldsymbol{\mathcal{D}}}_0 = {\boldsymbol{\mathcal{D}}}_1+ {\boldsymbol{\mathcal{D}}}_2 + 2\mathbb{E}\{(\check{\boldsymbol{\theta}}-{\boldsymbol{\theta}})(\hat{\boldsymbol{\theta}}-\check{\boldsymbol{\theta}})^T\},\nonumber
\end{equation}
or equivalently:
\begin{equation}\label{Distortion-UP-scalar}
\mathcal{D}_0 = \mathcal{D}_1+ \mathcal{D}_2 + 2\mathbb{E}\{(\check{\boldsymbol{\theta}}-{\boldsymbol{\theta}})^T(\hat{\boldsymbol{\theta}}-\check{\boldsymbol{\theta}})\},\nonumber
\end{equation}
where $\mathcal{D}_1\!=\!\tr({\boldsymbol{\mathcal{D}}}_1)$ and $\mathcal{D}_2\!=\!\tr({\boldsymbol{\mathcal{D}}}_2)$. Applying Cauchy-Schwarz inequality and using the fact $(x+y)^2 \! \leq \! 2(x^2+y^2)$ for $x,y \! \geq \!0$, we establish an upper bound on $\mathcal{D}_0$ as the following:
\begin{align}\label{Distortion-UP-scalar-2}
\mathcal{D}_0\leq (\sqrt{{\mathcal{D}}_1}+\sqrt{{\mathcal{D}}_2})^2\leq 2 ({\mathcal{D}}_1 +{\mathcal{D}}_2)=2{\mathcal{D}}.
\end{align}
Note that the upper bound on $\mathcal{D}_0$ in (\ref{Distortion-UP-scalar-2}) consists of two terms: the first term $2{\mathcal{D}}_1$ represents the MSE due to observation noises and quantization errors, whereas the second term $2\mathcal{D}_2$ is the MSE due to communication channel errors. In other words, {\it the contributions of observation noises and quantization errors in the upper bound are decoupled from those of communication channel errors}.

Relying on (\ref{Distortion-UP-scalar-2}), in the remaining of this section we derive two upper bounds on $\mathcal{D}$, denoted as ${\mathcal{D}}_a$ and ${\mathcal{D}}_b$, respectively, in sections \ref{D-a-derivations} and \ref{D-b-derivations},  in terms of $\{L_k, P_k\}_{k=1}^K$. While ${\mathcal{D}}_a$ is a tighter bound than  ${\mathcal{D}}_b$, its minimization demands a higher computational complexity. 
Leveraging on ${\mathcal{D}}_a$ and ${\mathcal{D}}_b$ expressions derived in this section, in sections \ref{optimization-algorithms} and \ref{optimization-algorithms-decoupled}, we propose two distinct schemes, which we refer to as ``coupled'' and ``decoupled'' schemes, to tackle the optimization problem formulated in (\ref{original Problem}), when ${\mathcal{D}}_0$ is replaced with ${\mathcal{D}}_a$ or ${\mathcal{D}}_b$.
%
\vspace{-0.3cm}
\subsection{Characterization of First Bound ${\mathcal{D}}_a$}\label{D-a-derivations}
Recall ${\mathcal{D}}\!=\!{\mathcal{D}}_1\!+\! {\mathcal{D}}_2$ in (\ref{Distortion-UP-scalar-2}). Below, we first derive ${\mathcal{D}}_1$. Deriving an exact expression for ${\mathcal{D}}_2$ remains elusive. Hence we derive an upper bound on ${\mathcal{D}}_2$, represented as ${\mathcal{D}}_2^{upb}$. Let ${\mathcal{D}}_a \!=\!{\mathcal{D}}_1\!+\! {\mathcal{D}}_2^{upb}$. Based on (\ref{Distortion-UP-scalar-2}), we have $\mathcal{D}_0\leq 2 \mathcal{D} \leq 2\mathcal{D}_a$. 


$\bullet$ {\bf Derivation of ${\mathcal{D}}_1$ in $\mathcal{D}_a$}: Since $\check{\boldsymbol{\theta}}$ is the linear MMSE estimator of $\boldsymbol{\theta}$ given $\boldsymbol{m}$, ${\boldsymbol{\mathcal{D}}_1}$  is the corresponding error covariance matrix. Consequently ${\mathcal{D}}_1\!=\!\tr({\boldsymbol{\mathcal{D}}}_1)$ is \cite{StevenKay1993}:
\begin{equation} \label{D1-formula}
{\mathcal{D}}_1=\tr(\mathcal{C}_{\boldsymbol{\theta}}-\mathbb{E}\{\boldsymbol{\theta}\boldsymbol{m}^T  \}(\mathbb{E}\{\boldsymbol{m}\boldsymbol{m}^T  \})^{-1} \mathbb{E}\{\boldsymbol{\theta}\boldsymbol{m}^T  \}^T),
\end{equation}
where $\mathbb{E}\{\boldsymbol{\theta}\boldsymbol{m}^T\}$ and $\mathbb{E}\{\boldsymbol{m}\boldsymbol{m}^T\}$, respectively, are cross-covariance and covariance matrices.  To find these matrices, we need to delve into statistics of quantization errors.
For sensor $k$, let the difference between observation $x_k$ and its quantization level $m_k$, i.e., $\epsilon_k\! =\! x_k-m_k$  be the corresponding quantization noise. In general, $\epsilon_k$'s are mutually correlated and also are correlated with $x_k$'s. However, in \cite{Widrow} 
it is shown that, when correlated Gaussian random variables are quantized with uniform quantizers of step sizes $\Delta_k$'s, quantization noises can be approximated as mutually independent random variables, that are uniformly distributed in the interval $[-\frac{\Delta_k}{2},\frac{\Delta_k}{2}]$, and are also independent of quantizer inputs. In this work, since $\boldsymbol{\theta}$ and $n_k$'s in (\ref{observation model}) are assumed Gaussian, $x_k$'s are correlated Gaussian that are quantized with uniform quantizers of quantization step sizes $\Delta_k$'s. Hence, $\epsilon_k$'s are approximated as mutually independent zero mean uniform random variables with variance $\sigma^2_{\epsilon_k}\!=\!\frac{\Delta^2_k}{12}$, that are also independent of $x_k$'s (and thus independent of $\boldsymbol{\theta}$ and $n_k$'s). These imply
$\mathbb{E}\{\boldsymbol{\theta}m_k  \}\!=\!\mathbb{E}\{\boldsymbol{\theta}(x_k-\epsilon_k)  \}\!=\!\mathcal{C}_{\boldsymbol{\theta}} \mathbf{a}_k$. Therefore:
\begin{equation}\label{theta m}
\mathbb{E}\{\boldsymbol{\theta}\boldsymbol{m}^T\}\!=\!\mathcal{C}_{\boldsymbol{\theta}}\mathbf{A}\!=\!\mathcal{C}_{{\boldsymbol{x}}{\boldsymbol{\theta}}}^T.
\end{equation}
Also, it is straightforward to verify:
\begin{equation}\label{mm^T}
\mathbb{E}\{\boldsymbol{m}\boldsymbol{m}^T\}\!=\! \mathbf{A}^T \mathcal{C}_{\boldsymbol{\theta}}\mathbf{A} \!+ \!\mathcal{C}_{\boldsymbol{n}}\!+\!\boldsymbol{Q}=\mathcal{C}_{\boldsymbol{x}}\!+\!\boldsymbol{Q},
\end{equation}
where $\boldsymbol{Q}\!=\!\mbox{diag}(\sigma^2_{\epsilon_1},\dots,\sigma^2_{\epsilon_K})$.
Substituting \eqref{theta m}, \eqref{mm^T} into \eqref{LMMSE-G-matrix}, \eqref{D1-formula} yield:
\begin{eqnarray}\label{G-definition}
\boldsymbol{G} \!& \! = \! &\! \mathcal{C}_{{\boldsymbol{x}}{\boldsymbol{\theta}}}^T(\mathcal{C}_{\boldsymbol{x}}\!+\!\boldsymbol{Q})^{-1}, \label{G-final-expression}\\ 
{\mathcal{D}}_1 \! & \! = \! & \! \tr(\mathcal{C}_{\boldsymbol{\theta}})\!- \!\tr(\mathcal{C}_{{\boldsymbol{x}}{\boldsymbol{\theta}}}^T (\mathcal{C}_{\boldsymbol{x}}\!+\!\boldsymbol{Q})^{-1}\mathcal{C}_{{\boldsymbol{x}}{\boldsymbol{\theta}}} ). \label{D1-final-expression}
\end{eqnarray}


$\bullet$ {\bf Derivation of ${\mathcal{D}}_2^{upb}$ in $\mathcal{D}_a$}:  Substituting $\check{\boldsymbol{\theta}}=\boldsymbol{G}\boldsymbol{m}$ and $\hat{\boldsymbol{\theta}}=\boldsymbol{G}\hat{\boldsymbol{m}}$ in $\boldsymbol{{\mathcal{D}}_2}$ we reach
$\boldsymbol{{\mathcal{D}}_2}\!=\!\boldsymbol{{{G}}}\boldsymbol{{\mathcal{M}}}\boldsymbol{{{G}^T}}$, where we define matrix $\boldsymbol{{\mathcal{M}}}\!=\!\mathbb{E}\{  (\hat{\boldsymbol{m}}-\boldsymbol{m})(\hat{\boldsymbol{m}}-\boldsymbol{m})^T\} $. Since  communication channel noises are mutually uncorrelated $\mathbb{E}\{(\hat{m}_i-m_i) (\hat{m}_j-m_j)\}\!=\!0~ \mbox{for} ~i\!\neq\!j$. Hence
$\boldsymbol{{\mathcal{M}}}$ is a diagonal matrix, whose $k$-th entry, $[\boldsymbol{{\mathcal{M}}}]_{k,k}\!=\!\mathbb{E}\{(\hat{m}_k-m_k)^2\}$, depends on the employed modulation scheme, channel gain ${\rvert h_k \rvert}$, channel noise variance $\sigma^2_{\omega_k}$, value of $\tau_k$, transmit power $P_k$, and quantization rate $L_k$. For our system model depicted in Section \ref{our-system-model}, in which sensors utilize BPSK modulation, we obtain (see Appendix \ref{upperboudn-m_prim-m-appnedix}):
\begin{equation}\label{bound-entries-M}
[\boldsymbol{{\mathcal{M}}}]_{k,k} \leq (\frac{4 \tau_k^2 L_k}{3}) \mbox{exp}(-\frac{\gamma_k P_k}{L_k})=u_k,
\end{equation} 
where $\gamma_k\!=\!\frac{   {{\rvert h_k \rvert}^2} }{ 2    {\sigma^2_{\omega_k}}}$ is  {\it channel to noise ratio} ($CNR$) for sensor $k$. Since $\boldsymbol{{\mathcal{M}}}$ and $\boldsymbol{{\mathcal{D}}_2}$ are semi-positive definite matrices, i.e., $\boldsymbol{{\mathcal{M}}} \! \succeq  \! 0, \boldsymbol{{\mathcal{D}}_2} \! \succeq \! 0$,  the bound in (\ref{bound-entries-M}) can provide us an upper bound on  ${\mathcal{D}}_2 \! = \! \tr(\boldsymbol{{\mathcal{D}}_2})$. Let $\boldsymbol{{\mathcal{M'}}}\!=\!\mbox{diag}({u}_1,...,{u}_K)$. Since  $\boldsymbol{{\mathcal{M'}}}  \! -\! \boldsymbol{{\mathcal{M}}} \! \succeq  \! 0$ and $ \boldsymbol{{{G}}}\boldsymbol{{\mathcal{M'}}}\boldsymbol{{{G}^T}}   \! \succeq  \! 0 $, we find \cite{vosoughi2006crb}:
\begin{equation}\label{D-2-upb}
{\mathcal{D}}_2 =\tr (\boldsymbol{{{G}}}\boldsymbol{{\mathcal{M}}}\boldsymbol{{{G}^T}}) \leq \tr(\boldsymbol{{{G}}}\boldsymbol{{\mathcal{M'}}}\boldsymbol{{{G}^T}})= {{\mathcal{D}}_2}^{upb},
\end{equation}
Regarding ${\mathcal{D}}_1$  and  ${\mathcal{D}}_{2}^{upb}$ a remark follows.

$\bullet$ {\bf Remark 1}: Note ${\mathcal{D}}_1$ in \eqref{D1-final-expression} 
only depends on quantization rates $L_k$'s, through the variances of quantization noises $\epsilon_k$'s in $\boldsymbol{Q}$. On the other hand,  ${\mathcal{D}}_2^{upb}$  in (\ref{D-2-upb}) depends on transmit powers $P_k$'s, through $\boldsymbol{{\mathcal{M'}}}$, as well as $L_k$'s through $\boldsymbol{{\mathcal{M'}}}$ and $\boldsymbol{Q}$ in $\boldsymbol{G}$. Hence, we derive the upper bound ${\mathcal{D}}_a\!=\!{\mathcal{D}}_1+ {\mathcal{D}}_2^{upb}$, in terms of the optimization parameters $\{L_k, P_k\}_{k=1}^K$.
%
\vspace{-0.3cm}
\subsection{Characterization of Second Bound ${\mathcal{D}}_b$}\label{D-b-derivations}

Recall from Section \ref{D-a-derivations} that ${\mathcal{D}}_a \! \!=\! \!{\mathcal{D}}_1\!+\! {\mathcal{D}}_2^{upb}$  where $\mathcal{D}_0 \! \leq \! 2 \mathcal{D} \! \leq \! 2\mathcal{D}_a$. Note that both ${\mathcal{D}}_1$ and  ${\mathcal{D}}_{2}^{upb}$ involve inversion of matrix $\mathcal{C}_{\boldsymbol{x}}\!+\!\boldsymbol{Q}$, incurring a high computational complexity that grows with $K$. 
For large $K$ such a matrix inversion, required to find the optimal resource allocation (see Section \ref{coupled-min-Da}), is burdensome. To curtail computational complexity, in this section we derive upper bounds on $\mathcal D_1$ and $\mathcal D^{upb}_2$,
represented as $\mathcal D_1^{upb}$ and $\mathcal D^{uupb}_2$, respectively, that do not involve such a matrix inversion.  
Let ${\mathcal{D}}_b \!=\!{\mathcal{D}}_1^{upb}\!+\! {\mathcal{D}}_2^{uupb}$. Based on (\ref{Distortion-UP-scalar-2}), we have $\mathcal{D}_0 \! \leq \! 2 \mathcal{D} \! \leq \! 2\mathcal{D}_a \! \leq \! 2\mathcal{D}_b $. 


$\bullet$ {\bf Derivation of ${\mathcal{D}}_1^{upb}$ in ${\mathcal{D}}_b$}: To find an upper bound on $\mathcal D_1$ in (\ref{D1-final-expression}) we use the following inequality \cite{Jun-Fang-2009}:

\begin{equation}\label{trace-inequality-general}
\tr (\bold E^T\bold F^{-1} \bold E)\geq \frac{(\tr (\bold E^T\bold E))^2}{\tr (\bold E^T\bold F \bold E)},
\end{equation}
where $\bold E$ is arbitrary and $\bold F  \! \succeq \! 0$. Recall $\boldsymbol Q$ is a diagonal matrix with non-negative entries, i.e., $\boldsymbol Q \! \succeq \! 0$. Also, $\mathcal{C}_{\boldsymbol{x}} \! \succeq \! 0$ since it is a covariance matrix. These imply $\boldsymbol Q\!+\!\mathcal{C}_{\boldsymbol{x}} \! \succeq \! 0$ \cite{vosoughi2006crb}. Applying  \eqref{trace-inequality-general} to (\ref{D1-final-expression}) we reach:
\begin{equation} \label{upperoundD1}
{\mathcal{D}}_1\!\leq\!\tr(\mathcal{C}_{\boldsymbol{\theta}}\!)-\frac{(\tr(\mathcal{C}_{{\boldsymbol{x}}{\boldsymbol{\theta}}}^T\mathcal{C}_{{\boldsymbol{x}}{\boldsymbol{\theta}}}))^2}{\tr(\mathcal{C}_{{\boldsymbol{x}}{\boldsymbol{\theta}}}^T (\mathcal{C}_{\boldsymbol{x}}\!+\!\boldsymbol{Q})\mathcal{C}_{{\boldsymbol{x}}{\boldsymbol{\theta}}})} = {\mathcal{D}}_1^{upb}.
\end{equation}
$\bullet$ {\bf Derivation of ${\mathcal{D}}_2^{uupb}$ in ${\mathcal{D}}_b$}: To find an upper bound on ${\mathcal{D}}_{2}^{upb}$ in (\ref{D-2-upb}), we 
take the following steps:
\begin{eqnarray} \label{trace(GMG)-bound}
{\mathcal{D}}_{2}^{upb} \overset{(a)}{=}  \sum_{k=1}^{K} \lambda_k (\boldsymbol G^T\boldsymbol G \boldsymbol {\mathcal M'} ) \overset{(b)}{\leq} \sum_{k=1}^{K} \lambda_k(\boldsymbol G^T\boldsymbol G) \lambda_k(\boldsymbol{\mathcal M'})\nonumber \\
\overset{(c)}{\leq} \lambda_{max}(\boldsymbol G^T\boldsymbol G) \sum_{k=1}^{K} \lambda_k(\boldsymbol{\mathcal M'}) =  \lambda_{max}(\boldsymbol G^T\boldsymbol G)\sum_{k=1}^{K} u_k,
\end{eqnarray}
where $(a)$ in (\ref{trace(GMG)-bound}) is obtained using the facts $\tr(\bold E \bold F)\! =\!\tr(\bold F \bold E)$ for arbitrary $\bold E, \bold F$ with matching sizes and $\tr(\bold E)\! =\! \sum_k \lambda_k (\bold E) \! =\! \sum_k [\bold E]_{k,k}$ for a square matrix $\bold E$ with eigenvalues $\lambda_k$'s
, $(b)$ is found using Theorem 9 in \cite {Fuzhen-Matrix-inequality},  and $(c)$ is true since  $\lambda_k (\boldsymbol G^T\boldsymbol G) \! \leq \! \lambda_{max}(\boldsymbol G^T\boldsymbol G)$ for $\forall k$.
%
Next, we derive an upper bound on $\lambda_{max}(\boldsymbol G^T\boldsymbol G)$ in (\ref{trace(GMG)-bound}). Using $\boldsymbol G$ in (\ref{G-final-expression}) we find $\boldsymbol G^T\boldsymbol G\!=\! (\mathcal{C}_{\boldsymbol{x}}\!+\!\boldsymbol{Q})^{-1}    \mathcal{C}_{{\boldsymbol{x}}{\boldsymbol{\theta}}}\mathcal{C}_{{\boldsymbol{x}}{\boldsymbol{\theta}}}^T    (\mathcal{C}_{\boldsymbol{x}}\!+\!\boldsymbol{Q})^{-1}$. 
Note $(\mathcal{C}_{\boldsymbol{x}}\!+\!\boldsymbol{Q})^{-1}$ and $\mathcal{C}_{{\boldsymbol{x}}{\boldsymbol{\theta}}}\mathcal{C}_{{\boldsymbol{x}}{\boldsymbol{\theta}}}^T$ are symmetric matrices. Therefore: 
\begin{eqnarray} \label{GGT-bound}
\!\!\!\!\!\! \lambda_{max} (\boldsymbol G^T\boldsymbol G)&\! \! \overset{(d)}{=} \! \!& ||\boldsymbol G^T\boldsymbol G||_2 \! \leq \!  [ || (\mathcal{C}_{\boldsymbol{x}}\!+\!\boldsymbol{Q})^{-1} ||_2]^2 || \mathcal{C}_{{\boldsymbol{x}}{\boldsymbol{\theta}}}\mathcal{C}_{{\boldsymbol{x}}{\boldsymbol{\theta}}}^T ||_2  \nonumber \\
&\! \! = \! \!& [\lambda_{max}((\mathcal{C}_{\boldsymbol{x}}\!+\!\boldsymbol{Q})^{-1})  ]^2 \lambda_{max} (\mathcal{C}_{{\boldsymbol{x}}{\boldsymbol{\theta}}}\mathcal{C}_{{\boldsymbol{x}}{\boldsymbol{\theta}}}^T)\nonumber \\
 &\! \! \overset{(e)}{=}  \! \!&[\frac{1}{\lambda_{min}(\mathcal{C}_{\boldsymbol{x}}\!+\!\boldsymbol{Q})} ]^2 \lambda_{max} (\mathcal{C}_{{\boldsymbol{x}}{\boldsymbol{\theta}}}\mathcal{C}_{{\boldsymbol{x}}{\boldsymbol{\theta}}}^T) \\
& \! \! \overset{(f)}{\leq} \! \! & [\frac{1}{\lambda_{min}(\mathcal{C}_{\boldsymbol{x}}\!)+   \underset {k}{\mbox{min}} (\sigma^2_{\epsilon_k})  } ]^2 \lambda_{max} (\mathcal{C}_{{\boldsymbol{x}}{\boldsymbol{\theta}}}\mathcal{C}_{{\boldsymbol{x}}{\boldsymbol{\theta}}}^T)=\tilde{\lambda} . \nonumber
\end{eqnarray}
$(d)$ holds due to the norm equality \cite{Rajendra}, $(e)$ is true since  $\lambda_{max}(\bold E^{-1})\!=\! \frac{1}{\lambda_{min}(\bold E)}$ for an invertible $\bold E$, and $(f)$ is obtained since for $\bold E,\bold F \! \succeq \! 0$ Weyl's inequality  \cite{Rajendra}  states
{$\lambda_{min} (\bold E + \bold F) \! \geq \! \lambda_{min} (\bold E) +\lambda_{min} (\bold F)$. Also, $\lambda_{min}(\bold Q)\! =\! \underset {k}{\mbox{min}} (\sigma^2_{\epsilon_k})$. Combining (\ref{trace(GMG)-bound}) and (\ref{GGT-bound}) we reach:
\begin{equation} \label{new-bound-D_2-upb-reserved}
{\mathcal{D}}_{2}^{upb} \! \leq \!    \tilde{\lambda} \sum_{k=1}^K u_k \!=\!{\mathcal{D}}_{2}^{uupb}.\nonumber
\end{equation}
Regarding ${\mathcal{D}}_1^{upb}$ and  ${\mathcal{D}}_{2}^{uupb}$ a remark follows.

$\bullet$ {\bf Remark 2}: Note ${\mathcal{D}}_1^{upb}$ in \eqref{upperoundD1} 
only depends on $L_k$'s, through the variances of quantization noises $\epsilon_k$'s in $\boldsymbol{Q}$. On the other hand,  ${\mathcal{D}}_2^{uupb}$  depends on $P_k$'s, through $u_k$'s, as well as $L_k$'s through $u_k$'s and $\sigma^2_{\epsilon_k}$'s in $\tilde{\lambda}$. Hence, we derive the upper bound ${\mathcal{D}}_b={\mathcal{D}}_1^{upb}+ {\mathcal{D}}_2^{uupb}$, in terms of $\{L_k, P_k\}_{k=1}^K$.
%
\vspace{-0.3cm}
\section{``Coupled'' Scheme For Resource Allocation}\label{optimization-algorithms}
So far, we have established $\mathcal{D}_0 \! \leq \! 2 \mathcal{D} \! \leq \!2\mathcal{D}_a \! \leq \! 2\mathcal{D}_b $, where $\mathcal{D}_a$ and $\mathcal{D}_b$ are derived in sections \ref{D-a-derivations} and \ref{D-b-derivations}, in terms of $\{L_k, P_k\}_{k=1}^K$. In this section we address the optimization problem formulated in (\ref{original Problem}), when ${\mathcal{D}}_0$ is replaced with ${\mathcal{D}}_a$ (in Section \ref{coupled-min-Da})
or ${\mathcal{D}}_b$ (in Section \ref{coupled-min-Db}). Note that (\ref{original Problem})  is a mixed integer nonlinear programming with exorbitant computational complexity \cite{NLIP-book-springer}. To simplify the problem, we temporarily relax the integer constraint on $L_k$'s and allow them to be positive numbers, i.e., we consider: 
\begin{eqnarray} \label{original Problem-version2}
 &&\underset{L_k,P_k, ~\forall k} {\text{minimize}} ~
 \mathcal{D}_0(\{L_k,P_k\}_{k=1}^K) \\
&& \text{s.t.}
\sum_{k=1}^K L_k\leq B_{tot}, \sum_{k=1}^K P_k\leq P_{tot}, L_k, P_k \in \mathbb{R}_{+}, ~\forall k. \nonumber
\end{eqnarray}
We propose ``coupled'' scheme to solve the relaxed problem in (\ref{original Problem-version2}), where the objective function is replaced with $\mathcal{D}_a$ or $\mathcal{D}_b$. In Section \ref{from-c-to-d} we discuss a novel approach to migrate from relaxed continuous $L_k$'s to integer $L_k$'s solutions.
\vspace{-0.3cm}
%
\subsection{Coupled Scheme for Minimizing $\mathcal{D}_a$}\label{coupled-min-Da}
 The quintessence of this ``coupled'' scheme follows.  We replace  $\mathcal D_0$ with $\mathcal D_a$  in (\ref{original Problem-version2}) and decompose the relaxed problem  into two sub-problems {\bf (SP1)} and {\bf (SP2)} as the following:
\begin{eqnarray} 
 \mbox{\bf (SP1)}&& \mbox{given} \{L_k\}_{k=1}^K, ~\underset{P_k, ~\forall k} {\text{minimize}} ~
 \mathcal{D}_a(\{P_k\}_{k=1}^K)\label{Power-optimization} \\
 && \text{s.t.}
\sum_{k=1}^K P_k\leq P_{tot},  ~P_k \in \mathbb{R}_{+}, ~\forall k, \nonumber\\
\mbox{\bf (SP2)}  && \mbox{given} \{P_k\}_{k=1}^K, ~\underset{L_k, ~\forall k} {\text{minimize}} ~
\mathcal{D}_a(\{L_k\}_{k=1}^K)  \label{quantization-bit-optimization}\\
&& \text{s.t.}
\sum_{k=1}^K L_k\leq B_{tot}, ~L_k \in \mathbb{R}_{+} , ~\forall k. \nonumber
 \end{eqnarray}
We iterate between solving these two sub-problems, until we reach the solution. 


$\bullet$ {\bf  Solving (SP1) Given in (\ref{Power-optimization})}: Considering Remark 1, we note that only $\mathcal{D}_2^{upb}$ in $\mathcal{D}_a$ depends on $P_k$'s. Hence, we replace the objective function in (\ref{Power-optimization}) with $\mathcal{D}_2^{upb}$. Since $\mathcal{D}_2^{upb}$ is a jointly convex function of $P_k$'s (see Appendix \ref{convexity-of-D^{upb}-wrs-P_k}) we use Lagrange multiplier method and solve the corresponding Karush-Kuhn-Tucker (KKT) conditions to find the solution. Substituting $u_k$ of
(\ref{bound-entries-M}) into $\boldsymbol{{\mathcal{M'}}}$ and noting that $\boldsymbol{G}$ does not depend on $P_k$'s, we rewrite  {\bf (SP1)} as below: 
\begin{eqnarray} \label{power-minimization-reformulate}
 &&\mbox{given} \{L_k\}_{k=1}^K, ~ \underset{P_k, ~\forall k}{\text{minimize}}
\sum_{k=1}^K {\alpha}_kL_k ~ \mbox{exp}(-     \frac    {\gamma_k P_k}{L_k}) ~\\
 &&~\text{s.t.}~
 \sum_{k=1}^K P_k\leq P_{tot}, ~
  P_k \in \mathbb{R}_{+}, ~\forall k, \nonumber
\end{eqnarray}
where ${\alpha}_k \!=\!(4\tau_k^2/3) || \boldsymbol g_k||^2 $ and $|| \boldsymbol g_k||^2$ is the squared Euclidean norm of the $k$-th column of $\boldsymbol{G}$.
Let $\mathcal{L}(\{L_k,P_k,\mu_k\}_{k=1}^K, \lambda)$ be the Lagrangian for  \eqref{power-minimization-reformulate}, where $\mu_k$ and $\lambda$ are the Lagrange multipliers. The corresponding KKT conditions are:
\begin{eqnarray} \label{KKT-power-Alg1}
&&\frac{\partial\mathcal{L}}{\partial P_k}=    -\alpha_k \gamma_k \mbox{exp}(-     \frac    {  \gamma_k P_k }{ L_k}) -\mu_k +\lambda=0, ~\forall k,
~\nonumber\\
&&P_k\mu_k=0, ~~\mu_k\geq 0, ~~ P_k\geq 0, ~\forall k, 
~\nonumber\\
&&\lambda(\sum_{k=1}^K P_k-P_{tot})=0, ~~\lambda\geq 0, ~~\sum_{k=1}^K P_k \leq P_{tot}. \nonumber
\end{eqnarray}
Since $\mathcal{D}_2^{upb}$ is a decreasing function of $P_k$'s and  $P_{tot}$ (see Appendix \ref{convexity-of-D^{upb}-wrs-P_k}), solving \eqref{power-minimization-reformulate} for $P_k$'s we find:

\begin{equation} \label{power-alo-alg1}
{P_k}=[\frac{L_k}{\gamma_k} \mbox{ln}(  \frac   { \gamma_k {\alpha}_k}{{\lambda}^*} )]^+,~~\forall k,~
\end{equation}
where $[x]^+=\mbox{max}(0,x)$,  $\sum_{k=1}^K P_k=P_{tot}$ and $\ln \lambda^*$ is:
%
%
%
\begin{equation} \label{Lagrange-multiplier-power-alo}
\ln {\lambda}^*={(\sum_{k \notin \mathcal I}  \frac {L_k}  { \gamma_k}  )^{-1}}[  -P_{tot} +  \sum_{k \notin \mathcal I} {\frac{L_k}{\gamma_k}} \ln(\gamma_k {\alpha}_k)   ].
\end{equation}
Set $\mathcal I \! = \! \{k \!:\! P_k \!= \!0,k\!=\!0,..., K\}$ in (\ref{Lagrange-multiplier-power-alo}) is the set of inactive sensors: sensors whose $L_k\!=\!0$ or $ \gamma_k {\alpha}_k \!< \! {\lambda}^*$, where $\gamma_k$ is $CNR$ of sensor $k$ and  ${\alpha}_k$ depends on the parameters of the observation model. Eq. (\ref{power-alo-alg1}) indicates that $P_k$ depends on both sensor observation and communication channel qualities, through $\gamma_k$  and $\alpha_k$. Also, sensor $k$ with a larger $L_k$ is allocated a larger $P_k$.
For asymptotic regime of large $P_{tot}$, we substitute \eqref{Lagrange-multiplier-power-alo} into \eqref{power-alo-alg1} and let $P_{tot} \to \infty$ to reach the following:
\begin{equation}\label{HP-power-allocation}
P_k=\frac{L_kP_{tot}}{\gamma_k \sum_{k \notin \mathcal I}\frac{L_k}{\gamma_k}},~~\forall k.
\end{equation}
Eq. (\ref{HP-power-allocation}) implies in this asymptotic regime, $P_k$ is proportional to $\frac{L_k}{\gamma_k}$. When $L_k$'s are equal, sensor with a smaller $CNR$ is allotted a larger $P_k$ (inverse of water-filling). When $\gamma_k$'s are equal, sensor with a larger $L_k$ is assigned a larger $P_k$. Regarding the solution in (\ref{power-alo-alg1}) two remarks follow.

$\bullet$ {\bf Remark 3}: For sensor $k$, we examine how $P_k$ varies as $\gamma_k$ changes, for a given $L_k$. We obtain:
\begin{equation}\label{power-alo-derivative}
\frac{\partial P_k}{\partial \gamma_k}=\frac{L_k}{\gamma_k^2} (1-\mbox{ln}(\frac{\gamma_k\alpha_k}{\lambda^*})),~~\forall k.
\end{equation}
Examining \eqref{power-alo-derivative} shows when $\gamma_k\alpha_k \! < \!e \lambda^*$, as $\gamma_k$ increases $P_k$ increases (water-filling). On the other hand, when $\gamma_k\alpha_k \! > \!e \lambda^*$, as $\gamma_k$ increases $P_k$ decreases (inverse of water-filling).

$\bullet$ {\bf Remark 4}:  For sensors $i,j$ we examine how $P_i,P_j$ are related to $\gamma_i,\gamma_j$. Suppose $L_i\!=\!L_j\!=\!L$ and $\alpha_i\!=\!\alpha_j\!=\!\alpha$. When $\frac{e \lambda^*}{\alpha} \!<\! \gamma_j \!<\! \gamma_i$ then $P_i \!<\! P_j$ (inverse of water-filling). On the other hand, when $\gamma_j \!<\! \gamma_i \!<\! \frac{e \lambda^* }{\alpha}$ then $P_i \!>\! P_j$ (water-filling).

%
$\bullet$ {\bf Solving (SP2) Given in (\ref{quantization-bit-optimization})}: 
Finding a closed-form solution for this problem remains elusive, due to non-linearity of the cost function and the fact that the inequality constraint on $L_k$'s  is not necessarily active. Let $\mathcal F\!=\!\{L_k \!:\! \sum_{k=1}^K L_k \! \leq \! B_{tot}, L_k \! \!\in \mathbb{R}_{+}, \forall k \nonumber\}$ be the feasible set of {\bf (SP2)}. To solve {\bf (SP2)} we use {\it a modified version of Ellipsoid method} \cite{stephen-boyd-lectures}. This cutting-plane optimization method is the generalized form of the one-dimensional bisection method for higher dimensions, and is theoretically efficient with guaranteed convergence \cite{stephen-boyd-lectures}.
The description of the method follows. Suppose the solution of {\bf (SP2)} is contained in an initial ellipsoid $\epsilon^0$ with center $\bold L'^{(0)}$ and shaped by matrix $\bold S^{(0)}\! \succeq \! 0$. The definition of $\epsilon^0$ is:
\begin{equation}\label{initial-ellipsoid}
\epsilon^0\! = \!\{\bold z \!: \! (\bold z-\bold L'^{(0)})^T {{\bold S}^{(0)}}^{-1} (\bold z-\bold L'^{(0)}) \!\leq \! 1\}.\nonumber
\end{equation}
For $\epsilon^0$ we choose a sphere that contains $\mathcal F$, with center $\bold L'^{(0)}\!=\!\frac{B_{tot}}{2}[1, ...,1]$ , radius $\frac{B_{tot}}{2}\sqrt K$, and thus $\bold S^{(0)}\!=\!(\frac{B_{tot}\sqrt{K}}{2}) \bold I_K$.
Essentially, this method uses gradient evaluation at iteration $i$, to discard half of $\epsilon^{i}$ and to form $\epsilon^{i+1}$ with center $\bold L'^{(i+1)}$, which is the minimum volume ellipsoid covering the remaining half of $\epsilon^{i}$. Note that $\epsilon^{i+1}$ can be larger than $\epsilon^{i}$ in diameter, however, it is proven that the volume of $\epsilon^{i+1}$ is smaller than that of $\epsilon^{i}$ and
center $\bold L'^{(i)}$ eventually converges to the solution of {\bf (SP2)}.   
To elaborate this method, suppose at iteration $i$, we have ellipsoid $\epsilon^i$ with center $\bold L'^{(i)}$ and shaped by matrix $\bold S^{(i)}$. Ellipsoid $\epsilon^{i+1}$ at iteration $i+1$ is obtained by evaluating the gradient $\nabla ^{(i)}$, defined below. When $\bold L'^{(i)} \! \in \! \mathcal F$ then $\nabla^{(i)}$ is the gradient of the objective function (so-called objective cut) evaluated at $\bold L'^{(i)}$. When $\bold L'^{(i)} \! \notin \! \mathcal F$ then $\nabla^{(i)}$ is the gradient of the inequality constraint that is being violated (so-called feasibility cut) evaluated at $\bold L'^{(i)}$. The update steps are:
\begin{eqnarray}\label{half-plane-ellipsoid}
\epsilon^{i+1}  \!\!&\!\!=\!\!&\!\! \epsilon^{i} \cap \{\bold z \!: \! \nabla^{(i)T} (\bold z-\bold L'^{(i)} )\leq 0\},\nonumber\\
\bold L'^{(i+1)}\!\!&\!\!=\!\!&\!\! \bold L'^{(i)}-\frac{1}{K+1} \bold S^{(i)} \tilde \nabla ^{(i)}, ~~ \tilde \nabla ^{(i)}\!=\! \frac{\nabla ^{(i)}}{\sqrt{ \nabla ^{(i)T} \bold S^{(i)} \nabla ^{(i)}}}, \label{ellipsoid-update} \nonumber\\
\bold S^{(i+1)} \!\!&\!\!=\!\!&\!\!\frac{K^2}{K^2-1} (\bold S^{(i)}-\frac{2}{K+1}\bold S^{(i)}\tilde \nabla ^{(i)}   \tilde \nabla ^{(i)T} \bold S^{(i)} ),\nonumber
\end{eqnarray}
in which:
\[ \nabla ^{(i)}= \begin{cases}
    \nabla^{(i)}_{oc} ,& \text{if } ~~~~~\sum_{k=1}^K L'^{(i)}_k\leq B_{tot}, ~ L'^{(i)}_k \in \mathbb{R}_{+}, ~\forall k, \nonumber\\
    \nabla^{(i)}_{sfc},            & \text{if}  ~~~~~\sum_{k=1}^K L'^{(i)}_k >  B_{tot}, ~L'^{(i)}_k   \in \mathbb{R}_{+}, ~\forall k,   \\
    \nabla^{(i)}_{nfc}                & \text{if}  ~~~~~L'^{(i)}_j \le 0, ~~ \mbox{for~some}~j \in\{1, ..., K\},  \nonumber
\end{cases}\]\label{cases-gradient}
where  $\nabla^{(i)}_{oc}$ ,$\nabla^{(i)}_{sfc}$ and $\nabla^{(i)}_{nfc}$ are objective cut, rate-sum constraint feasibility cut and nonnegative rate feasibility cut, respectively, evaluated at $\bold L'^{(i)}$:
\begin{eqnarray}
\!\!\!\!\!\!\!\!\!\! \nabla^{(i)}_{oc} \!\!&\!\!=\!\!&\!\! [\frac{\partial{\mathcal{D}_a}}{\partial{L_1}}|_{L_1\!=\!L'^{(i)}_1}, ...,\frac{\partial{\mathcal{D}_a}}{\partial{L_K}}|_{L_K\!=\!L'^{(i)}_K}]^T, \label{gradient-oc} \\
  \!\!\!\!\!\!\!\!\!\! \nabla^{(i)}_{sfc} \!\!&\!\!=\!\!&\!\!  [\frac{\partial{{(\sum_{j=1}^K L_j)}}}{\partial{L_1}}|_{L_1\!=\!L'^{(i)}_1}, ..., \frac{\partial{{(\sum_{j=1}^K L_j)}}}{\partial{L_K}}|_{L_K\!=\!L'^{(i)}_K}]^T, \label{gradient-sfc} \\
 \!\!\!\!\!\!\!\!\!\!   \nabla_{nfc}^{(i)} \!\!&\!\!=\!\!&\!\!  [\frac{-\partial{{L_j}}}{\partial{L_1}}|_{L_1\!=\!L'^{(i)}_1}, ..., \frac{-\partial{{L_j}}}{\partial{L_K}}|_{L_K\!=\!L'^{(i)}_K}]^T. \label{gradient-nfc} 
 \end{eqnarray}
After some mathematical manipulations and using the fact $\frac{\partial \bold E^{-1}}{\partial x}=-\bold E^{-1} \frac{\partial \bold E}{\partial x}\bold E^{-1}$, we find $\frac{\partial{\mathcal{D}_a}}{\partial{L_k}}~\forall k$ in (\ref{gradient-oc})  is equal to below:
\begin{eqnarray} \label{derivative-objective1-to-L-ver2}
\!\!\!\! \tr( \boldsymbol G [\frac{\partial \boldsymbol Q}{\partial L_k}\! -\! \frac{\partial \boldsymbol Q}{\partial L_k}(\mathcal{C}_{\boldsymbol{x}}\!+\! \boldsymbol{Q})^{-1}\boldsymbol {\mathcal M'}
\!+\!\frac{\partial {\boldsymbol {\mathcal M'}}}{\partial L_k} \! - \! \boldsymbol{\mathcal M'} (\mathcal{C}_{\boldsymbol{x}}\!+\!\boldsymbol{Q})^{-1}\frac{\partial \boldsymbol {Q}}{\partial L_k}] \boldsymbol G^T),\nonumber
\end{eqnarray}
in which $\frac{\partial \boldsymbol Q}{\partial L_k}$ and $\frac{\partial \boldsymbol{\mathcal M'}}{\partial L_k}$ are all-zero matrices, except for one non-zero element in each matrix $[\frac{\partial \boldsymbol Q}{\partial L_k}]_{k,k}\!=\!\frac {\partial \sigma^2_{\epsilon_k} }{\partial L_k}\!=\!- \frac {2 \ln 2~ \tau_k^2 ~2^{L_k} }{3(2^{L_k}-1)^3}$ and
$[\frac{\partial \boldsymbol {\mathcal M'}}{\partial L_k}]_{k,k}\!=\!\frac{\partial u_k}{\partial L_k}\!=\!\frac{4 \tau_k^2}{3}\mbox{exp}(\frac{- \gamma_k P_k }{ L_k})[1+ \frac{\gamma_k P_k}{L_k}]$.
Furthermore,  $\nabla^{(i)}_{sfc}$ in (\ref{gradient-sfc}) is a vector of all ones and $\nabla^{(i)}_{nfc}$  in (\ref{gradient-nfc}) is a vector of all zeros and $-1$ for its $j$-th entry. 
%
%
%
%
\begin{figure*}
  \centering
  
  \subcaptionbox{$\boldsymbol {L}'^{(i)} \! \in \! \mathcal F \! \Rightarrow \! \nabla^{(i)}\!=\!\nabla^{(i)}_{oc}$, update $\epsilon^{i}$ with center  $\boldsymbol {L}'^{(i)}$ to $\epsilon^{i+1}$ with center  $\boldsymbol {L}'^{(i+1)}$ \label{as}}[5.7cm]{\vspace{-.2cm}\includegraphics[width=1.4in]{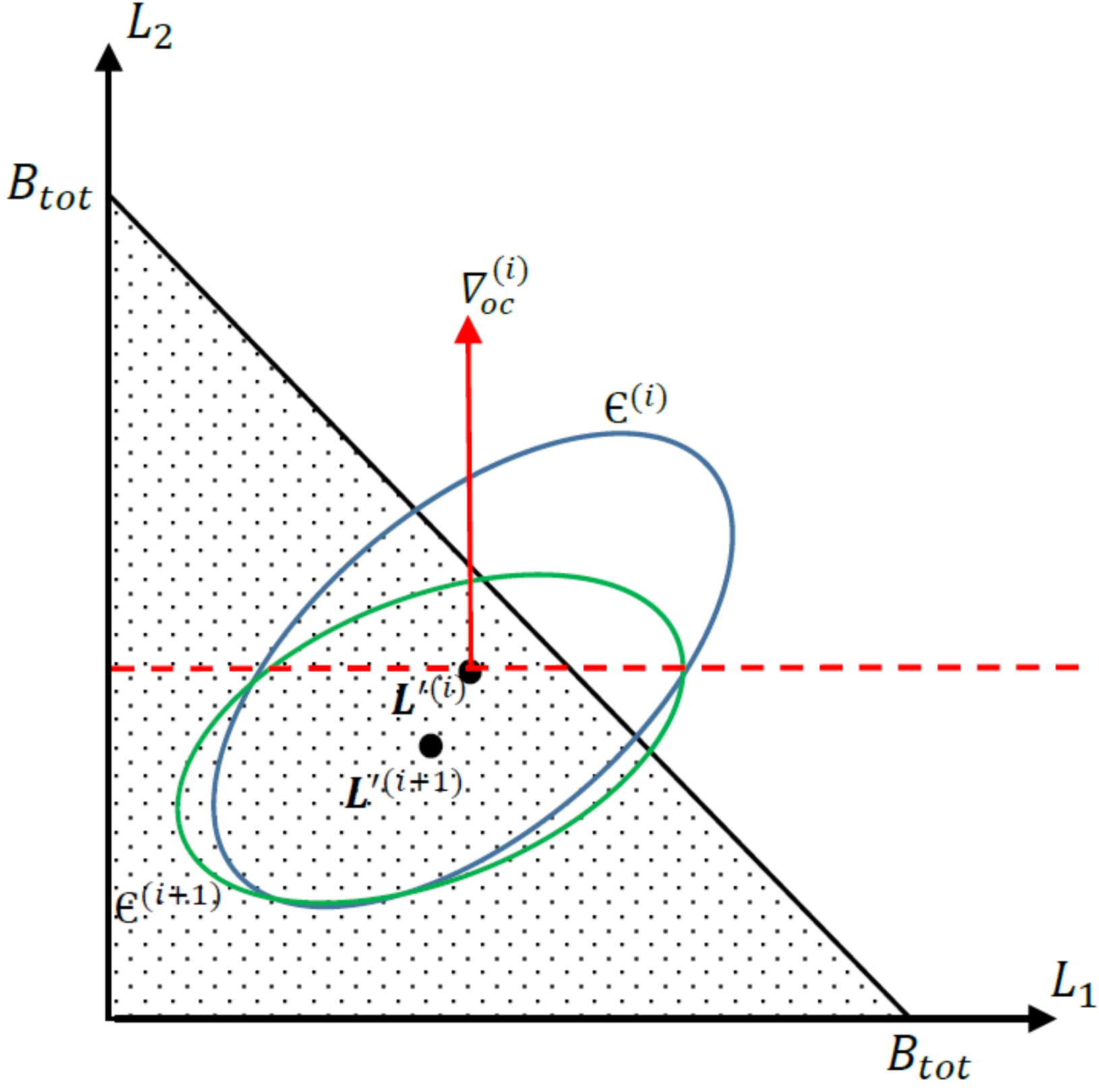}}\hfill%
  \subcaptionbox{$\boldsymbol {L}'^{(i)} \! \notin \! \mathcal F$  since $L'^{(i)}_1+L'^{(i)}_2 \! > \! B_{tot} \Rightarrow \nabla^{(i)}\!=\!\nabla^{(i)}_{sfc}$, update $\epsilon^{i}$ with center  $\boldsymbol {L}'^{(i)}$ to $\epsilon^{i+1}$ with center  $\boldsymbol {L}'^{(i+1)}$\label{fig1:b}}[5.7cm]{\vspace{-.2cm}\includegraphics[width=1.35in]{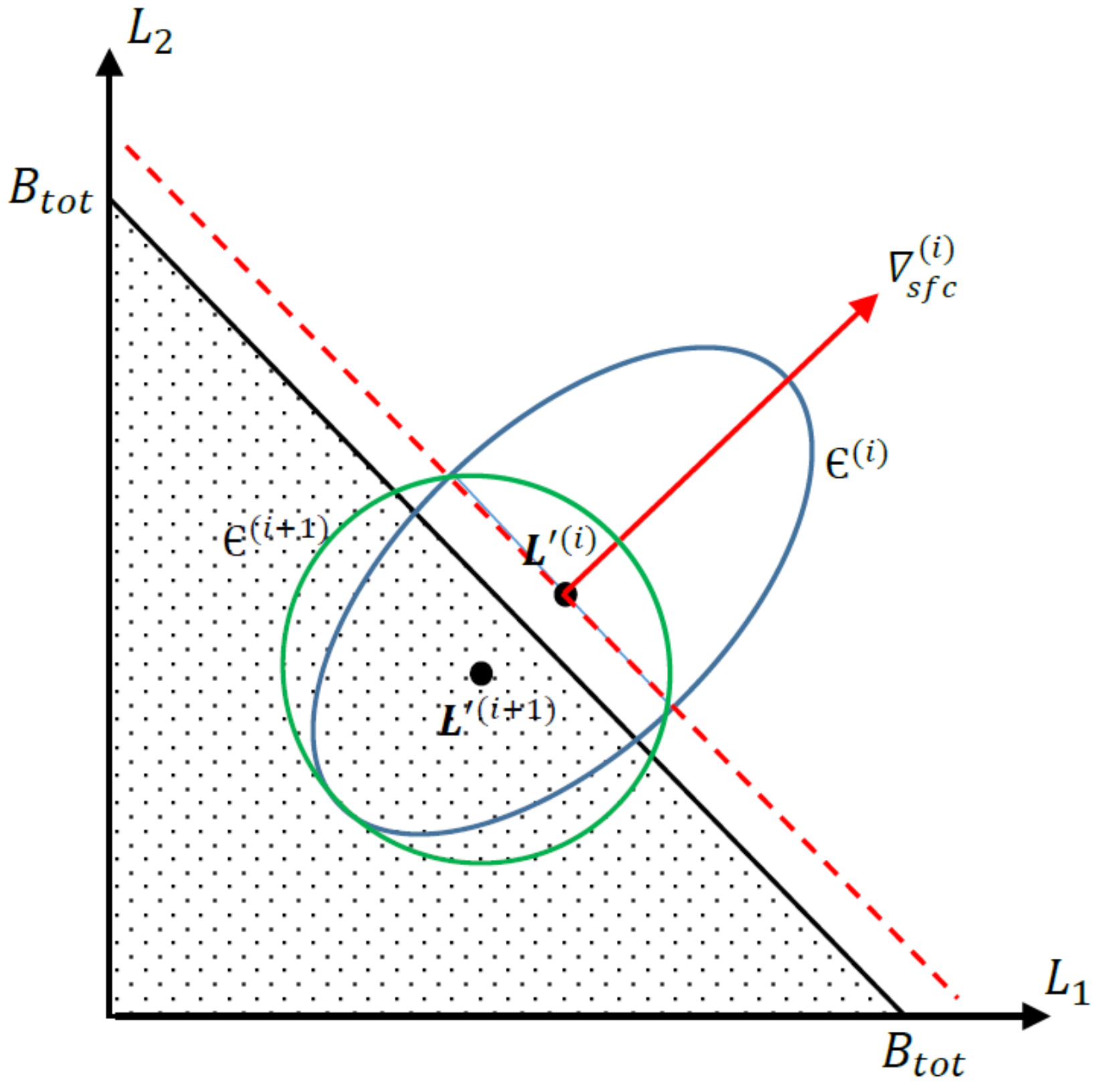}}\hfill%
\subcaptionbox{$\boldsymbol {L}'^{(i)}  \! \notin \! \mathcal F$ since $L'^{(i)}_1\! < \! 0 \! \Rightarrow \! \nabla^{(i)}\!=\!\nabla^{(i)}_{nfc}$, update $\epsilon^{i}$ with center  $\boldsymbol {L}'^{(i)}$ to $\epsilon^{i+1}$ with center  $\boldsymbol {L}'^{(i+1)}$\label{fig1:b}}[5.7cm]{\vspace{-.2cm}\includegraphics[width=2in]{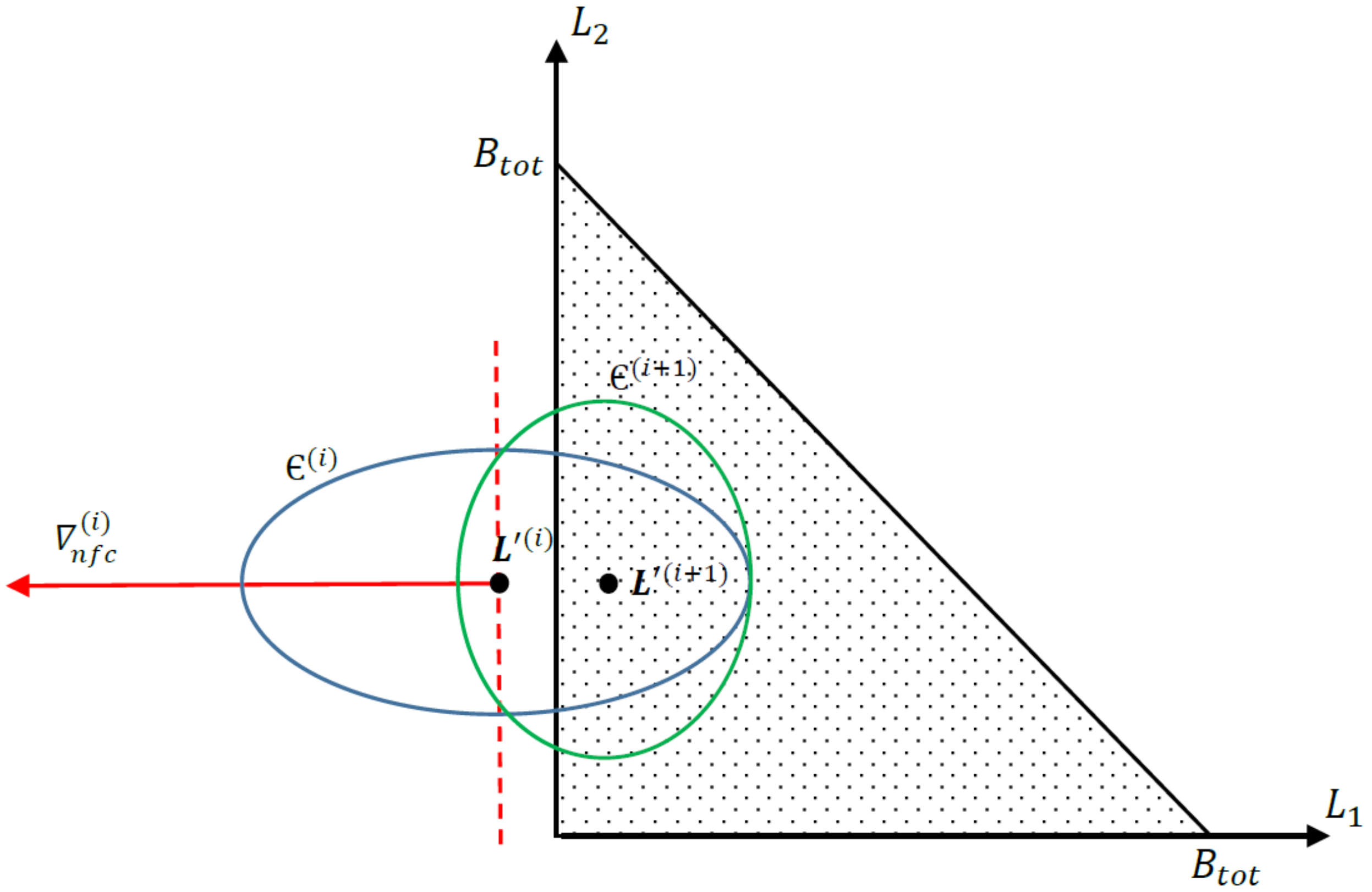}}\hfill%
\vspace{-.2cm}
\caption{Modified ellipsoid method for constrained optimization porblem}
\vspace{-.6cm}
\label{Fig1-ellipsoid}
\end{figure*}
%
%
%
\begin{algorithm}
 \KwData{System parameters defined in Section \ref{our-system-model}}
 \KwResult{Continuous solutions for optimization parameters   $\bold L^c\!=\![L^c_1,...,L^c_K],   \bold P^c\!=\![P^c_1,...,P^c_K]$}
  initialization\;
$j=0$,
$~\bold L^{c(0)}\! =\! \frac{B_{tot}}{2}[1, ...,1], ~\bold P^{c(0)}\! =\! [0, ...,0]$ \\
  \While{$1$}{
  \small $\boldsymbol{G}^{(j)}\!\!=\!\mathcal{C}_{{\boldsymbol{x}}{\boldsymbol{\theta}}}^T(\mathcal{C}_{\boldsymbol{x}}\!+\!\mbox{diag}(\frac{\tau_1^2}{3(2^{L_1^{(i)}}-1)^2},...,\frac{\tau_K^2}{3(2^{L_K^{(j)}}-1)^2}))^{-1}$\\
$  {\alpha}_k^{(j)} \!=\!(4\tau_k^2/3) || \boldsymbol g_k^{(j)}||^2$,~~$\forall k$\\ 
$   P^{c(j)}_k\!\!=\![\frac{L^{c(j)}_k}{\gamma_k} \mbox{ln}(  \frac   { \gamma_k {\alpha}^{(j)}_k}{e^{-P_{tot}} \prod_{k \notin \mathcal I} (\gamma_k {\alpha}_k^{(j)})^{\frac{L_k^{(j)}}{\gamma_k}}   ]^{(\sum_{k \notin \mathcal I}  \frac {L_k^{(j)}}  { \gamma_k}  )^{-1}}} )]^+$\\
 \vspace{.1cm}
\hbox to \hsize{\dashfill\hfil}
 \begin{center}
 \text{Modified ellipsoid method}
 \vspace{-.3cm}
 \end{center}
 \hbox to \hsize{\dashfill\hfil}
modified ellipsoid method initialization\;
$i=0$,
$~\bold L'^{(0)}\!=\!\frac{B_{tot}}{2}[1, ...,1]$,~ $\bold S^{(0)}\!=\!(\frac{B_{tot}\sqrt{K}}{2}) \bold I $\\
%
\While{$1$}{
$\bold L'^{(i+1)}=\bold L'^{(i)}-\frac{1}{K+1} \bold S^{(i)} \tilde \nabla ^{(i)}$\\
$\bold S^{(i+1)}=\frac{K^2}{K^2-1} (\bold S^{(i)}-\frac{2}{K+1}\bold S^{(i)}\tilde \nabla ^{(i)}   \tilde \nabla ^{(j)T} \bold S^{(i)} )$\\
$i=i+1$

\If{$\sqrt {\nabla^{(i+1)T} \bold S^{(i+1)} \nabla^{(i+1)} }<\varepsilon$ ~$\vee$ ~$i> I_{max}$}{break}
}
\hbox to \hsize{\dashfill\hfil}
$~\bold L^{c(j+1)}=\bold L'^{(i+1)}$\\
$j=j+1$\\
\If {$\mathcal{D}_a(\bold P^{c(j)},\bold L^{c(j)})-\mathcal{D}_a(\bold P^{c(j-1)},\bold L^{c(j-1)})<\eta$ ~$\vee$~ $j>J_{max}$}{break} 
 }
\TitleOfAlgo{``$a$-coupled'' algorithm for minimizing $\mathcal{D}_a$}
\vspace{-1.0cm}
 \end{algorithm}
%
As stopping criterion, we check if $\sqrt {\nabla^{(i)T} \bold S^{(i)} \nabla^{(i)} }<\varepsilon$, where $\varepsilon$ is a predetermined error threshold, or if the number of iterations exceeds a predetermined maximum $I_{max}$. 
Fig. \ref{Fig1-ellipsoid} illustrates the above ellipsoid method for $K\!=\!2$ sensors, where the feasible set $\mathcal F$ is the triangle with the hatch pattern. 

%
In nutshell, we have analytically solved {\bf (SP1)} and provided an iterative ellipsoid method with guaranteed convergence to address {\bf (SP2)}. With these ammunitions we tackle the problem in (\ref{original Problem-version2}), when $\mathcal{D}_0$ is replaced with $\mathcal{D}_a$. Let vectors $\bold L^c\!=\![L_1^c,...,L_K^c], \bold P^c\!=\![P^c_1,...,P^c_K]$ denote the solutions to (\ref{original Problem-version2}). We take an iterative approach that switches between solving {\bf (SP1)} and {\bf (SP2)},  until we converge to vectors $\bold L^c$ and $\bold P^c$. 
As stopping criterion, we check if the decrease in $\mathcal{D}_a$ in two consecutive iterations is less than a predetermined error threshold $\eta$, or if the number of switching between solving {\bf (SP1)} and {\bf (SP2)} exceeds a predetermined maximum $J_{max}$.  The ``$a$-coupled'' algorithm summarizes the steps described above, guaranteeing  that $\mathcal{D}_a$ decreases in each iteration $j$. Regarding ``$a$-coupled'' algorithm a remark follows.

$\bullet$ {\bf Remark 5}: Implementing the ellipsoid method requires a $K$-dimensional search. Also, in inner loop where $\boldsymbol {L}'^{(i)} \! \in \! \mathcal F$ we have $\nabla^{(i)}\!=\!\nabla^{(i)}_{oc}$, which needs inversion of $\mathcal{C}_{\boldsymbol{x}}\!+\!\boldsymbol{Q}$. In outer loop inversion of $\mathcal{C}_{\boldsymbol{x}}\!+\!\boldsymbol{Q}$ is required to update $\alpha_k^{(j)}$.
\vspace{-0.3cm}
%
\subsection{Migration from Continuous to Integer Solutions for Rates}\label{from-c-to-d}
We describe an approach on how to migrate from continuous solution $\bold L^c$ to integer solution $\bold L^d$. Let $\mathcal X$ be the index set of sensors which quantization rates are discretized. Initially $\mathcal X$ is an empty set.
We consider two scenarios: (i) $\sum_{j  \in \mathcal X} L^d_j \! + \! \sum_{j  \notin \mathcal X} L^c_j \!< \! B_{tot}$, (ii)  $\sum_{j  \in \mathcal X} L^d_j \! + \! \sum_{j  \notin \mathcal X} L^c_j \! = \! B_{tot}$.  
When case (i) occurs, it means that minimizing $\mathcal{D}_a$ has not been negatively impacted by bandwidth shortage.
Hence, we discritize the rate of sensor $j$ with the smallest $L_j^c$, since this sensor is more likely to be the weakest player in the network, in the sense that it has the minimal contribution to $\mathcal{D}_a$. We round $L_j^c$ ``up'' or ``down'', depending on which one yields a smaller $\mathcal{D}_a$ and consider the discritized $L_j^d$ as fixed and final value. 
When case (ii) occurs, it is very likely that minimizing $\mathcal{D}_a$  has been negatively impacted by bandwidth shortage and some sensors were imposed smaller rates, compared with an unlimited $B_{tot}$ scenario. Note that in case (ii), rounding up the rate of any sensor, would enforce decreasing the rates of some other sensors. Hence, we should discritize in a way that, the positive impact of rounding up a rate on  $\mathcal{D}_a$ would dominate the negative effect of decreasing the rates of some other sensors on $\mathcal{D}_a$.
Therefore, we discritize the rate of sensor $j$ with the largest $L_j^c$, since this sensor is more likely to be the strongest player, in the sense that it has the maximal contribution to $\mathcal{D}_a$. We round $L_j^c$ ``up'' and consider the discritized $L_j^d$ as fixed and final value. 
After each discritization, we need to update  $\mathcal X$ and the available bandwidth to $B_{tot} \!- \!\sum_{j \in \mathcal X} L_j^d$, and apply ``$a$-coupled'' algorithm to reallocate $P_{tot}$ among all sensors and $B_{tot} \!- \!\sum_{j \in \mathcal X} L_j^d$ among those sensors with continuous valued rates. We continue this procedure  until $\mathcal X$ includes all sensors.
\vspace{-0.7cm}
%
\subsection{Coupled Scheme for Minimizing $\mathcal{D}_b$}\label{coupled-min-Db}
Similar to Section \ref{coupled-min-Da}, in this section we consider two sub-problems, which we refer to as $(\bf{SP3})$ and $(\bf{SP4})$. They are similar to 
\eqref{Power-optimization} and \eqref{quantization-bit-optimization}, with the difference that $\mathcal D_a$ is replaced with $\mathcal D_b$. Solutions to $(\bf{SP3})$ and $(\bf{SP4})$ follow.

$\bullet$ {\bf  Solving (SP3)}:
Considering Remark 2, we note that only $\mathcal{D}_2^{uupb}$ in $\mathcal{D}_b$ depends on $P_k$'s. Hence, we replace the objective function in {(\bf {SP3})} with $\mathcal{D}_2^{uupb}$. Since $\mathcal{D}_2^{uupb}$ is a jointly convex function of $P_k$'s (see Appendix \ref{convexity-of-D^{upb}-wrs-P_k}) we use Lagrange multiplier method and solve the corresponding KKT conditions to find the solution.
Also, $\mathcal{D}_2^{uupb}$ is a decreasing function of $P_k$'s and  $P_{tot}$ (see Appendix \ref{convexity-of-D^{upb}-wrs-P_k}). Therefore, solving {\bf (SP3)} for $P_k$'s we find:
\begin{eqnarray} 
{P_k} \!\!&\!\!=\!\!&\!\! [\frac{L_k}{\gamma_k} \mbox{ln}(  \frac   { \gamma_k {\tau_k^2}}{{\mu}^*} )]^+, ~\forall k,~ \mbox{where}~\sum_{k=1}^K P_k=P_{tot},\label{power-alo-alg2}\\
\ln {\mu}^* \!\!&\!\!=\!\!&\!\! {(\underset{k \notin \mathcal J}\sum  \frac {L_k}  { \gamma_k})^{-1}}[  -P_{tot} +\underset{k \notin \mathcal J}\sum{ \frac{L_k}{\gamma_k} }  \ln (\gamma_k \tau_k^2)   ] . \label{Lagrange-multiplier-power-ALG2}
\end{eqnarray}
%
%
Set $\mathcal J \! = \! \{k \!:\! P_k \!= \!0,k\!=\!0,..., K\}$ in (\ref{Lagrange-multiplier-power-ALG2}) is the set of inactive sensors: sensors whose $L_k\!=\!0$ or $ \gamma_k {\tau}_k^2 \!< \! {\mu}^*$. 
It is noteworthy to mention that for asymptotic regime of large $P_{tot}$, we have the same power allocation policy as in \eqref{HP-power-allocation}.

$\bullet$ {\bf  Solving (SP4)}:  We apply the modified ellipsoid method we used for {\bf (SP2)}, to solve {\bf (SP4)}. The feasible set $\mathcal F$ and the update steps are similar, with a difference: the gradient of the objective function $\nabla^{(i)}_{oc}$ changes and instead of $\frac{\partial{\mathcal{D}_a}}{\partial{L_k}}$, $\frac{\partial{\mathcal{D}_b}}{\partial{L_k}}$ needs to be derived. 
However, $\frac{\partial{\mathcal{D}_b}}{\partial{L_k}}\!=\!\frac{\partial{\mathcal{D}_1^{upb}}}{\partial{L_k}}\!+\!\frac{\partial{\mathcal{D}_2^{uupb}}}{\partial{L_k}}$, where:
 \begin{eqnarray}
  \frac{\partial{\mathcal{D}_1^{upb}}}{\partial{L_k}}=\frac{-(\tr(\mathcal{C}_{{\boldsymbol{x}}{\boldsymbol{\theta}}}^T\mathcal{C}_{{\boldsymbol{x}}{\boldsymbol{\theta}}}))^2\tr(\mathcal{C}_{{\boldsymbol{x}}{\boldsymbol{\theta}}}^T  \frac{\partial \bold Q}{\partial L_k} \mathcal{C}_{{\boldsymbol{x}}{\boldsymbol{\theta}}})}{(\tr(\mathcal{C}_{{\boldsymbol{x}}{\boldsymbol{\theta}}}^T (\mathcal{C}_{\boldsymbol{x}}\!+\!\boldsymbol{Q})\mathcal{C}_{{\boldsymbol{x}}{\boldsymbol{\theta}}}))^2},\nonumber
  \end{eqnarray}
%
%
\begin{align} \label{D_2^{uupb}- derivative-L_j}
 \frac{\partial{\mathcal{D}_2^{uupb}}}{\partial{L_k}}=\begin{cases}
    \tilde \lambda~ \frac{\partial u_k}{\partial L_k} ,& \text{if } k\notin \mathcal A, \nonumber\\
    \tilde \lambda~ [\frac{\partial u_k}{\partial L_k}-\frac{2\frac{\partial \sigma^2_{\epsilon_k}}{\partial L_k} \sum_{j=1}^K u_j}{\lambda_{min} ( {\mathcal C_x})+\sigma^2_{\epsilon_k}}], & \text{if}~ k \in \mathcal A, \nonumber
\end{cases}
\end{align}
in which \textcolor{black}{ set $\mathcal A \! = \! \{k \!:\! k \!=\!\underset{i} {\mbox{argmin}}(\sigma^2_{\epsilon_i})\}$}, $\tilde \lambda$ is defined in (\ref{GGT-bound}) and $ \frac{\partial \bold Q}{\partial L_k}, \frac{\partial u_k}{\partial L_k}$ are given  in Section \ref{coupled-min-Da}.
Having the solutions for $(\bf{SP3})$ and $(\bf{SP4})$, we can address the problem in (\ref{original Problem-version2}), when $\mathcal{D}_0$ is replaced with $\mathcal{D}_b$, utilizing an iterative algorithm similar to ``$a$-coupled'' algorithm 
outlined in Section \ref{coupled-min-Da}, which we call ``$b$-coupled'' algorithm, and a discretizing approach similar to Section \ref{from-c-to-d}. Different from  ``$a$-coupled'' algorithm, ``$b$-coupled'' algorithm does not require matrix inversion, although implementing the modified ellipsoid method still requires a $K$-dimensional search. 
\vspace{-0.3cm}
%
\section{``Decoupled'' Scheme For Resource Allocation}\label{optimization-algorithms-decoupled}
In Section  \ref{optimization-algorithms} we proposed two iterative coupled schemes, which minimize $\mathcal D_a$ and $\mathcal D_b$. In both schemes we resorted to the iterative modified ellipsoid method to find $L_k$'s, since finding a closed-form solution for $L_k$'s remained elusive.
We recall the discussion at the beginning of Section \ref{section-characterization-MSE}, which indicates $\mathcal D_1$ (and its bound  $\mathcal D_1^{upb}$) represent the MSE due to observation noises and quantization errors, whereas  $\mathcal D_2$ (and its bounds  $\mathcal D_2^{upb}, \mathcal D_2^{uupb}$) are the MSE due to communication channel errors. 
Leveraging on this decoupling of the contributions of observation noises and quantization errors from those of communication channel errors, we propose ``decoupled'' scheme  to minimize these decoupled contributions separately, and to find the optimization parameters $\{L_k,P_k\}_{k=1}^K$ in closed-form expressions, eliminating the computational burden of the modified ellipsoid method for conducting $K$-dimensional search and finding $\bold L^c$ vector. Similar to Section \ref{optimization-algorithms} we start with ``decoupled'' scheme to solve the relaxed problem,  via allowing $L_k$'s to be positive numbers. 
\vspace{-0.3cm}
%
\subsection { Decoupled Scheme for Minimizing $\mathcal D_a$}
The essence of ``decoupled'' scheme is to solve the following two sub-problems in a sequential order:
\begin{eqnarray} 
\!\!\!\!\!\!\!\!\!\!\!\!\!\!\!\!\!\!\!\!\!\!\!\!\!\!\!\!\!\! \mbox{\bf (SP5)} && \underset{L_k, ~\forall k} {\text{minimize}} ~
  \mathcal{D}_1^{upb}(\{L_k\}_{k=1}^K) \label{quantization-rate-optimization-Alg3} \\
  && \text{s.t.}
 \sum_{k=1}^K L_k\leq B_{tot}, ~L_k \in \mathbb{R}_{+} , ~\forall k, \nonumber\\
\!\!\!\!\!\! \mbox{\bf (SP6)} && \mbox{given $\{L_k\}_{k=1}^K$},  ~\underset{P_k, ~\forall k} {\text{minimize}} ~
   \mathcal{D}_2^{upb}(\{P_k\}_{k=1}^K) \label{Power-optimization-Alg3} \\
   && \text{s.t.}
    \sum_{k=1}^K P_k\leq P_{tot},  ~P_k \in \mathbb{R}_{+}, ~\forall k. \nonumber
  \end{eqnarray}
Different from Section  \ref{optimization-algorithms} there is no iteration between  {\bf (SP5)} and {\bf (SP6)}.
After solving  {\bf (SP5)} and {\bf (SP6)}, we take a similar approach to Section \ref{from-c-to-d}, to migrate from continuous to integer solutions for $L_k$'s.

$\bullet$ {\bf  Solving (SP5)}: Considering \eqref{upperoundD1}, we realize that minimizing $\mathcal{D}_1^{upb}$ is equivalent to minimizing 
${\tr(\mathcal{C}_{{\boldsymbol{x}}{\boldsymbol{\theta}}}^T \boldsymbol{Q}\mathcal{C}_{{\boldsymbol{x}}{\boldsymbol{\theta}}})} \!= \! \sum_{k=1}^{K} \delta_k \sigma^2_{\epsilon_k}$, where $\delta_k$
%
is the squared Euclidean norm of  the $k$-th row of $\mathcal{C}_{{\boldsymbol{x}}{\boldsymbol{\theta}}}$. 
Since $\sum_{k=1}^{K} \delta_k \sigma^2_{\epsilon_k}$ is a jointly convex function of $L_k$'s (See Appendix \ref{proof-convexity-D_1^{upb}}) we use Lagrange multiplier method and solve the corresponding KKT conditions to find the solution. Also, $\sum_{k=1}^{K} \delta_k \sigma^2_{\epsilon_k}$ is a decreasing function of $L_k$'s and $B_{tot}$ (See Appendix \ref{proof-convexity-D_1^{upb}}). Therefore, solving {\bf (SP5)} in \eqref{quantization-rate-optimization-Alg3} for $L_k$'s and using the approximation $2^{L_k} \!- \!1 \! \approx \! 2^{L_k}$ we reach:
\begin{eqnarray}
\!\!\!\!\!\!\! L_k \!\!&\!\!= \!\!&\!\! [\log_2(\sqrt{\frac{\tau^2_k\delta_k 2 \ln 2}{3\nu^* }})]^+, ~\forall k \label{L_i-solution-d1^upb, minimization}~ \mbox{where}~ \sum_{k=1}^K L_k\!=\! B_{tot},\\
\!\!\!\!\!\!\! \nu^* \!\!&\!\!= \!\!&\!\!\frac{2\ln 2}{3} [{ 4^{-B_{tot}}\prod^{K}_{k=1} \delta_k \tau^2_k}]^{\frac{1}{K}}.\label{nu-solution-d1^upb, minimization}
\end{eqnarray}
Substituting \eqref{nu-solution-d1^upb, minimization} into \eqref{L_i-solution-d1^upb, minimization} one can verify:
\begin{equation} \label{L_i-solution-final}
{L_k}\! =\!  \left[\frac{B_{tot}}{K}   +  {\log}_2\left(  \sqrt{\frac   {\delta_k \tau^2_k}   { (\prod_{i=1}^K\delta_i \tau^2_i)^{\frac{1}{K}}} } \right)\right]^+, ~\forall k.
\end{equation}
Examining (\ref{L_i-solution-final}), we note that the first term inside the bracket is common among all sensors, whereas the second term differs among sensors and depends on $\delta_k, \tau_k$. 
For $\tau_k\!=\! \tau, ~\forall k$, 
sensor $k$ with a larger $\delta_k$ (i.e., better observation quality) is allocated a larger $L_k$. 
Note that, since $\mathcal D_1^{upb}$ does not capture communication channel errors, $L_k$ in (\ref{L_i-solution-final}) is independent of sensor communication channel quality. This is different from solution of  {\bf (SP2)} in (\ref{quantization-bit-optimization}), where $L_k$'s depend on both sensor observation and communication channel qualities. 
 
$\bullet$ {\bf  Solving (SP6)}: Similar to {\bf (SP1)} in \eqref{Power-optimization}, the objective function in (\ref{Power-optimization-Alg3}) is a jointly convex function of $P_k$'s and also decreases as $P_k$'s and $P_{tot}$ increase (See appendix \ref{convexity-of-D^{upb}-wrs-P_k}). Indeed, solving  {\bf (SP6)} yields the exact solution as that of {\bf (SP1)}, provided in (\ref{power-alo-alg1}) and \eqref{Lagrange-multiplier-power-alo}.

$\bullet$ {\bf Remark 6}: We minimize ${\mathcal{D}}^{upb}_1$ in {\bf (SP5)} instead of $\mathcal D_1$, since it yields a closed form solution for $L_k$'s given in (\ref{L_i-solution-final}). In Section \ref{deplete-B_tot} we discuss minimizing $\mathcal D_a \!= \! \mathcal D_1\!+\! \mathcal D_2^{upb}$, when
we substitute \eqref{L_i-solution-final} into $\mathcal D_1$ and (\ref{power-alo-alg1}), \eqref{Lagrange-multiplier-power-alo}, \eqref{L_i-solution-final}  into  $\mathcal D_2^{upb}$.
\vspace{-0.3cm}
%
\subsection{Does Depleting $B_{tot}$ always reduce $\mathcal D_a$?}\label{deplete-B_tot}
To answer this question, we consider the solution in \eqref{L_i-solution-final} for asymptotic regime of large $B_{tot}$. As $B_{tot} \! \rightarrow \! \infty$, we find $ L_k \! \rightarrow \! \frac{B_{tot}}{K}, \forall k$, i.e., we should equally distribute $B_{tot}$ among sensors. 
In situations where $B_{tot}$ is large and communication channel quality of sensor $k$ is poor (i.e., small $\gamma_k$ due to small channel gain $|h_k|$ or small $P_k$ due to low $P_{tot}$), the solution in \eqref{L_i-solution-final} can lead to a large value for $\mathcal D_a$, since sending a large number of bits $L_k$ over poor quality channels increases communication channel errors and thus  $ \mathcal{D}_2^{upb}$.
This observation suggests that perhaps, we should first find $B^{opt}$ (for $B^{opt} \!< \!B_{tot}$), where $B^{opt}$ depends on channel gains and $P_{tot}$, and then distribute $B^{opt}$ among sensors, to control the growth of $\mathcal{D}_2^{upb}$ in $\mathcal D_a$. 
In fact, when we substitute \eqref{L_i-solution-final} into $\mathcal D_1$ and (\ref{power-alo-alg1}), \eqref{Lagrange-multiplier-power-alo}, \eqref{L_i-solution-final} into  $\mathcal D_2^{upb}$, we find $\mathcal D_1$ is a decreasing function of $B_{tot}$, whereas $\mathcal D_2^{upb}$ is an increasing function of $B_{tot}$. As $B_{tot}\! \rightarrow \! 0$, $\mathcal D_1$ approaches its maximum value $\tr (\mathcal C_{\boldsymbol \theta})$, i.e., trace of covariance of unknowns, whereas $\mathcal D_2^{upb}$ goes to zero.  On the other hand, as $B_{tot} \! \rightarrow \! \infty$, $\mathcal D_1$ approaches its minimum value $d_0 \! = \! \tr(\mathcal C_{\bold \theta})-\tr(\mathcal C_{\boldsymbol x \bold \theta }^T \mathcal C_{\boldsymbol x}^{-1} \mathcal C_{\boldsymbol x \bold \theta })$, i.e., clairvoyant centralized estimation where unquantized sensor observations are available at the FC, whereas $\mathcal D_2^{upb}$ grows unboundedly  (See Appendix \ref{properties-on-distortion-elements}). 
This tradeoff suggests that there should be a value $B^{opt}$ that minimizes $\mathcal D_a$. %
\begin{figure}[t]
   \includegraphics[width=3.5in]{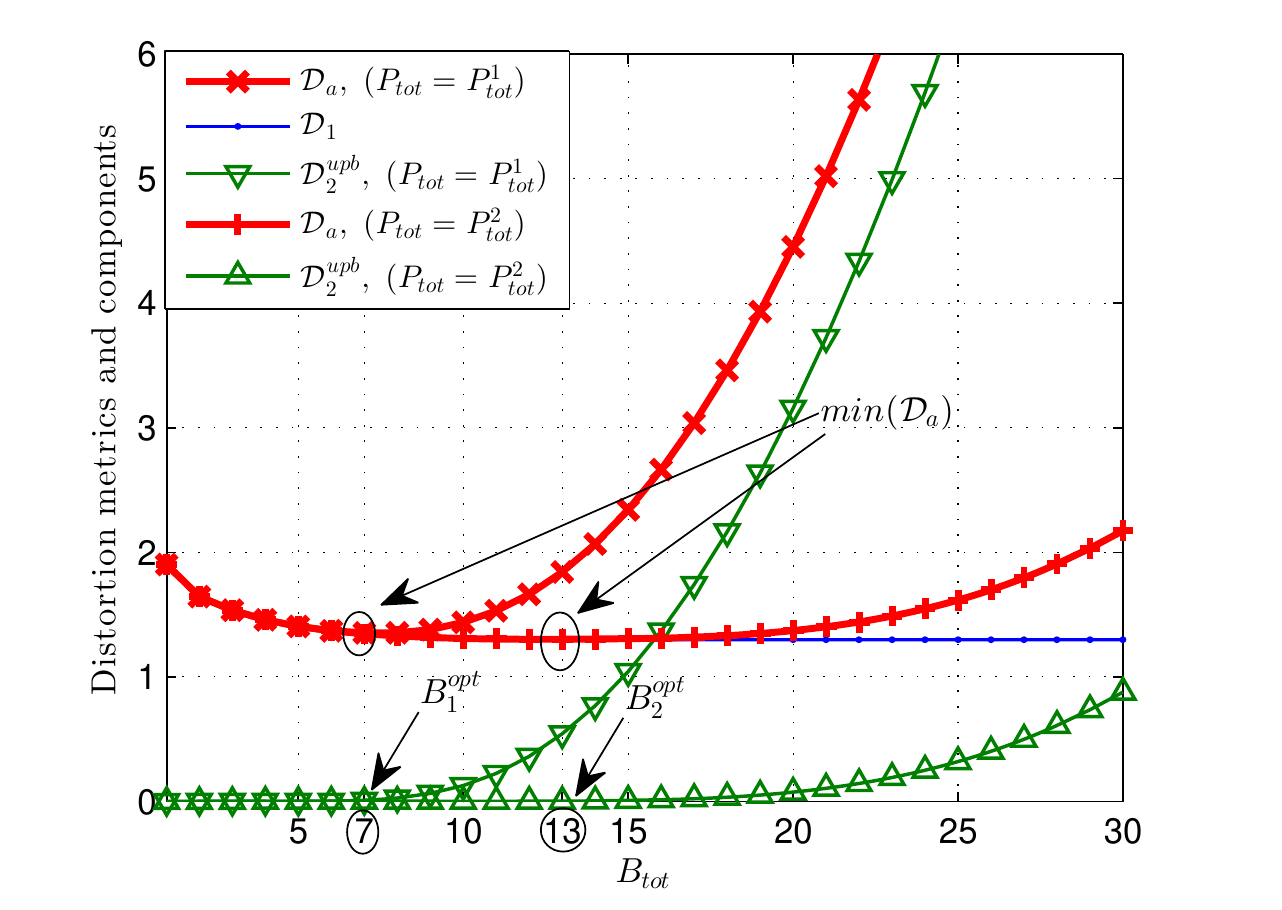}
   \vspace{-0.75cm}
   \caption{$\mathcal D_1, \mathcal D_2^{upb}, \mathcal D_a$ versus $B_{tot}$ (measured in bits)}
   \label{justification-decoupled-1}
   \vspace{-0.55cm}
   \end{figure}
%
Fig. \ref{justification-decoupled-1} illustrates  $\mathcal D_1, \mathcal D_2^{upb}, \mathcal D_a$ versus $B_{tot}$
for two values $P_{tot}^1,P_{tot}^2$, where $P_{tot}^2\geq P_{tot}^1$. Fig. \ref{justification-decoupled-1} shows that $B_2^{opt} \! \geq \!B_1^{opt}$. This observation can be explained as the following.
%
%
Note that $\mathcal D_1$ is independent of $P_{tot}$, whereas  $\mathcal D_2^{upb}$ decreases as $P_{tot}$ increases (Appendix \ref{convexity-of-D^{upb}-wrs-P_k} shows $\frac{\partial \mathcal D_2^{upb}}{P_{k}} \! \leq \!0~ \forall k$, and thus $\frac{\partial \mathcal D_2^{upb}}{P_{tot}} \! \leq \! 0$). 
Hence, as $P_{tot}$ increases we can transmit a larger number of bits, i.e., larger $B^{opt}$, without incurring an increase in communication channel errors.
%
%

Motivated by these, we propose ``$a$-decoupled'' algorithm. This algorithm starts with initiating $B^{opt}\!=\!1$ and increasing the value of $B^{opt}$ by one bit at each iteration, where the maximum number of iterations is $B_{tot}$. At iteration $i$, we find $L_k^{(i)}$'s using 
\eqref{L_i-solution-final} and $P_k^{(i)}$'s using (\ref{power-alo-alg1}),\eqref{Lagrange-multiplier-power-alo}, substitute these $L_k^{(i)},P_k^{(i)}$'s into 
$\mathcal D_a$, and check if the decrease in $\mathcal{D}_a$ in two consecutive iterations is less than zero, i.e., $\mathcal{D}_a(\{L_k^{(i)},P_k^{(i)}\}_{k=1}^K)-\mathcal{D}_a(\{L_k^{(i-1)},P_k^{(i-1)}\}_{k=1}^K) \! < \! 0$.  
If this inequality holds for $i \! < \!B_{tot}$, we let $B^{opt}\!=\!i$ and find the solution for $L_k$'s using 
\eqref{L_i-solution-final} and $P_k$'s using (\ref{power-alo-alg1}), \eqref{Lagrange-multiplier-power-alo}, when $B_{tot}$ is substituted with $B^{opt}$. Otherwise, we let $B^{opt}\!=B_{tot}$ and find the solution for $L_k$'s using 
\eqref{L_i-solution-final} and $P_k$'s using (\ref{power-alo-alg1}),  \eqref{Lagrange-multiplier-power-alo}. At the end, we take the approach in Section \ref{from-c-to-d} to migrate from continuous to integer solutions for rates and find the corresponding powers. 
\begin{figure}[t]
\includegraphics[width=3.3in]{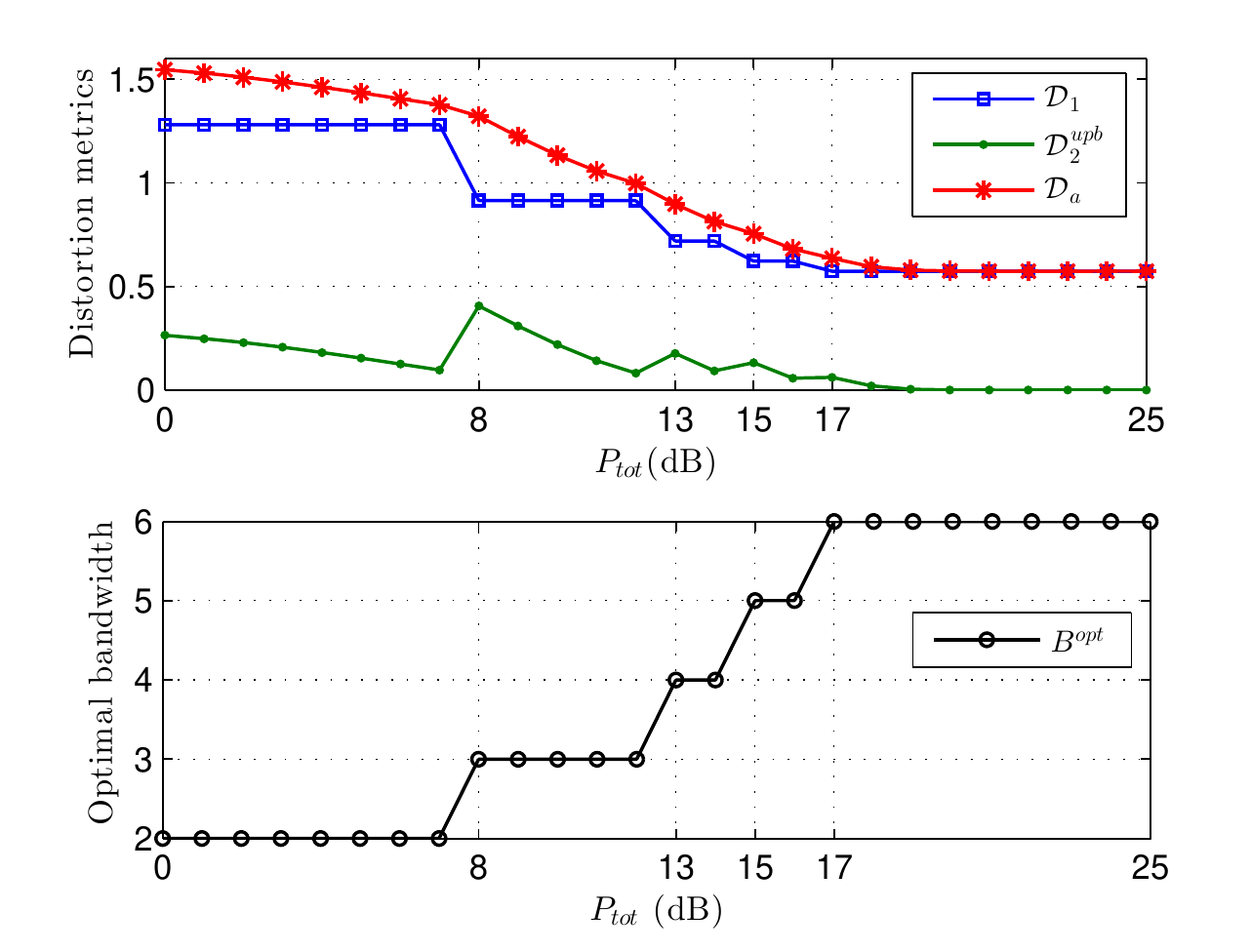}
\vspace{-0.2cm}
\caption{behavior of ``$a$-decoupled'' algorithm}
\label{decoupled-algorithm-justification-2}
\vspace{-0.6cm}
 \end{figure}
%
Fig. \ref{decoupled-algorithm-justification-2} illustrates the behavior of ``$a$-decoupled'' algorithm and in particular, how $B^{opt}, \mathcal D_1, \mathcal D_2^{upb}, \mathcal D_a$ vary as $P_{tot}$ increases. Note that as $P_{tot}$ increases $B^{opt}$  remains constant for certain (long) intervals and increases for some other (short) intervals of $P_{tot}$ values. 
The behavior of $B^{opt}$ versus $P_{tot}$ dictates the behavior of $\mathcal D_1, \mathcal D_2^{upb}, \mathcal D_a$ as well. 
For the $P_{tot}$ intervals where $B^{opt}$ is fixed, $\mathcal D_1$ is fixed as well, since it is independent of $P_{tot}$, whereas $\mathcal D_2^{upb}$ and thus $\mathcal D_a$ decrease as $P_{tot}$ increases. For the  $P_{tot}$ intervals where $B^{opt}$ increases, $\mathcal D_1$ decreases and $\mathcal D_2^{upb}$ increases (Appendix \ref{properties-on-distortion-elements} shows $\frac{\partial{\mathcal{D}_1}}{\partial{B_{tot}}} \leq 0$ and $\frac{\partial{\mathcal{D}_2^{upb}}}{\partial{B_{tot}}} \geq 0$). Overall, as $P_{tot}$ increases $\mathcal D_a$  decreases, since the decrease in $\mathcal D_1$, when $B^{opt}$ increases, dominates the increase in $\mathcal D_2^{upb}$.
%
%
%
%
%
Regarding the modified ``$a$-decoupled'' algorithm a remark follows.

$\bullet$ {\bf Remark 7}:  The modified ``$a$-decoupled'' algorithm requires only one dimensional search to find $B^{opt}$ and thus $L_k$'s \eqref{L_i-solution-final}, as opposed to $K$-dimensional search required by the modified ellipsoid method in ``$a$-coupled'' algorithm. Also, finding $B^{opt}$ at most needs $B_{tot}$ number of iterations and no switching between solving {\bf (SP5)} and {\bf (SP6)} is needed.  
\vspace{-0.3cm}
\subsection {Decoupled Scheme for Minimizing $\mathcal D_b$ }
Note that ``$a$-decoupled'' algorithm still requires inversion of matrix $\mathcal C_{\boldsymbol x }+ \boldsymbol Q$ to calculate $\alpha_k$ and find $P_k$ using (\ref{power-alo-alg1}), \eqref{Lagrange-multiplier-power-alo}. 
To eliminate this matrix inversion, we propose to minimize $\mathcal D_2^{uupb}$ instead of $\mathcal D_2^{upb}$.
%
%
Since $\mathcal D_2^{uupb}$ is a jointly convex function of $P_k$'s  (see Appendix \ref{convexity-of-D^{upb}-wrs-P_k}),  substituting $L_k$'s of  \eqref{L_i-solution-final} into $\mathcal D_2^{uupb}$ and minimizing it with respect to $P_k$'s, we reach the solution provided in (\ref{power-alo-alg2}), (\ref{Lagrange-multiplier-power-ALG2}). Let  ``$b$-decoupled'' algorithm be the one that minimizes $\mathcal D_1^{upb}$ and $ \mathcal D_2^{uupb}$ separately in $\mathcal D_b$. This algorithm is very similar to  ``$a$-decoupled'' algorithm described in Section \ref{deplete-B_tot}, with the difference that when finding $B^{opt}$, at iteration $i$ we find $L_k^{(i)}$'s using 
\eqref{L_i-solution-final} and $P_k^{(i)}$'s using (\ref{power-alo-alg2}), (\ref{Lagrange-multiplier-power-ALG2}), substitute these $L_k^{(i)},P_k^{(i)}$'s into 
$\mathcal D_b$, and check if the decrease in $\mathcal{D}_b$ in two consecutive iterations is less than zero. After finding 
 $B^{opt}$ we find the solution for $L_k$'s using \eqref{L_i-solution-final} and $P_k$'s using 
\eqref{power-alo-alg2}, (\ref{Lagrange-multiplier-power-ALG2}).
The behavior of ``$b$-decoupled'' algorithm and in particular, how $B^{opt}, \mathcal D_1^{upb}, \mathcal D_2^{uupb}, \mathcal D_b$ vary as $P_{tot}$ increases, is similar to ``$a$-decoupled'' algorithm depicted in Fig. \ref{decoupled-algorithm-justification-2} and is omitted due to lack of space.
\vspace{-0.3cm}
%
\section{Numerical and Simulation Results}\label{smulation-section}
In this section through simulations we corroborate our analytical results. Without loss of generality and for the sake of illustration we set $K\!=\!3$,  $\mathcal C_{\boldsymbol{\theta}}\!=\![1~ (\sqrt{2}/2);(\sqrt{2}/2)~ 2]$, \textcolor{black}{ $\textbf {a}_1\!=\![1~ 1]^T$, $\textbf{a}_2\!=\![0.6~ 0.6]^T$, $\textbf{a}_3\!=\![0.4~ 0.4]^T$,} $\sigma^2_{n_k}\!=\!1$, $\sigma^2_{w_k}\!=\!1$, $h_k\!=\!1$, for all algorithms. ``CS" and ``DS" in the legends of the figures indicate the continuous solutions and discrete solutions of the algorithms, respectively. 

%
Figs. \ref{power-vs-power-coupled-1} and \ref{rate-vs-power-Coupled-1} illustrate $\{10\log_{10} (P_k)\}_{k=1}^3$ and $\{L_k\}_{k=1}^K$ vs. $P_{tot}$, respectively, for ``$a$-coupled'' algorithm and $B_{tot}\!=\!30, 3$ bits. 
Figs.\ref{power-vs-power-coupled-1}.a and \ref{rate-vs-power-Coupled-1}.a for $B_{tot}\!=\!30$ bits (abundant bandwidth) show that as $P_{tot}$ increases, both power and rate allocation approaches uniform allocation. This is in agreement with \eqref{HP-power-allocation}. When $P_{tot}$ is small, only sensor 1 is active. As $P_{tot}$ increases sensors 2 and 3 become active in sequential order.
Figs. \ref{power-vs-power-coupled-1}.b and \ref{rate-vs-power-Coupled-1}.b for $B_{tot}\!=\!3$ bits (scarce bandwidth) show that only sensor 1 is active and $P_1\!=\!P_{tot}$.
Overall, these observations imply that power and rate allocation depends on both $P_{tot}$ and $B_{tot}$, e.g., when we have plentiful $P_{tot}$ and scarce $B_{tot}$ only the sensor with the largest observation gain is active.
Also, uniform power and rate allocation is near optimal when we have ample $P_{tot}$ and $B_{tot}$.\\
Figs. \ref{power-vs-power-decoupled-1} and \ref{rate-allocation-A-decoupled-1} depict $\{10\log_{10} (P_k)\}_{k=1}^3$ and $\{L_k\}_{k=1}^K$ vs. $P_{tot}$, respectively, for ``$a$-decoupled'' algorithm and $B_{tot}\!=\!30, 3$ bits. 
Similar to ``$a$-coupled'' algorithm, we observe that when both $P_{tot}$ and $B_{tot}$ are abundant, uniform power and rate allocation is near optimal. This is in agreement with \eqref{HP-power-allocation}, \eqref{L_i-solution-final}. On the other hand, when $B_{tot}$ is scarce, power and rate allocation is way different from being uniform. In fact, when $P_{tot}$ is ample, \eqref{HP-power-allocation}  indicates that $P_k$
%
is proportional to $L_k$. Also, when $B_{tot}$ is scarce,  \eqref{L_i-solution-final} states that $L_k$'s and consequently  $P_k$'s  are not uniformly distributed. These indicate that power and rate allocation is affected by sensors' observation qualities and channel gains, as well as both $P_{tot}$ and $B_{tot}$. 
There are two slight differences between ``$a$-coupled'' and ``$a$-decoupled'' algorithms:
%
(i) for $B_{tot}\!=\!30$ bits (ample bandwidth) sensors 2 and 3 become active at smaller $P_{tot}$ values in ``a-decoupled'' algorithm, (ii) for $B_{tot}\!=\!3$ bits (scarce bandwidth), ``$a$-decoupled'' algorithm ultimately activates all sensors at $P_{tot}$ increases, while  ``$a$-coupled'' algorithm only activates sensor 1. Note that for scarce bandwidth, even when $P_{tot}$ is very large, power and rate allocation in ``$a$-decoupled'' algorithm is non-uniform.\\
%
%
Figs. \ref{power-vs-BW-coupled-1} and \ref{rate-vs_BW-Coupled-1} depict $\{10\log_{10} (P_k)\}_{k=1}^3$ and $\{L_k\}_{k=1}^K$ vs. $B_{tot}$, respectively, for ``$a$-coupled'' algorithm and $P_{tot}\!=\!16, 30$ dB. 
The observations in these figures are in full agreement with the former ones. In particular, for $P_{tot} \!=\!30$ dB (large power), when $B_{tot}$ is small, only sensor 1 is active. As $B_{tot}$ increases, sensors 2 and 3 become active too, in a way that power and rate allocation approaches to uniform for large  $B_{tot}$. On the other hand, for $P_{tot} \!=\!16$ dB only sensor 1 is active and $P_1\!=\!P_{tot}$.\\
%
%
%
Figs. \ref{power-vs_BW-decoupled-1} and  \ref{rate-vs_BW-decoupled-1}  illustrate $\{10\log_{10} (P_k)\}_{k=1}^3$ and $\{L_k\}_{k=1}^K$ vs. $B_{tot}$, respectively, for ``$a$-decoupled'' algorithm and $P_{tot}\!=\!16, 30$ dB. 
While the behaviors of ``$a$-coupled'' and ``$a$-decoupled'' have similarities, they have differences as the following: (i) for $P_{tot}\!=\!30$ dB sensors 2 and 3 become active at smaller $B_{tot}$ value in ``$a$-decoupled'' algorithm,  (ii) for $P_{tot}\!=\!16$ dB, ``$a$-decoupled'' algorithm ultimately activates all sensors as $B_{tot}$ increases, while  ``$a$-coupled'' algorithm only activates sensor 1. 
%
Note for $P_{tot} \!=\!16$ dB, even when $B_{tot}$ is very large, power and rate allocation in ``$a$-decoupled'' algorithm is non-uniform.
this is because $B^{opt} \! <\! B_{tot}$ in this case and according to \eqref{L_i-solution-final}, (\ref{power-alo-alg1}), \eqref{Lagrange-multiplier-power-alo} power and rate allocation would be non-uniform.

%
In Fig. \ref{MSE-upperbound-comparison} we plot $\mathcal D_a$ when implementing ``$a$-coupled'' and ``$a$-decoupled''  and $\mathcal D_b$ when implementing ``$b$-coupled'' and ``$b$-decoupled'' algorithms,   vs. $P_{tot}$, for $B_{tot}\!=\! 30, 3$ bits. 
We observe that ``$a$-coupled'' and ``$b$-decoupled''  algorithms perform the best and the worst, respectively. Also, All the algorithms outperform uniform resource allocation (except for ``$b$-decoupled'' algorithm when $B_{tot} \!=\!3$ bits and $13 \!<\!P_{tot} \!<\!18$ dB). 
For small $P_{tot}$, the performance of each algorithm does not change as we decrease $B_{tot} \!=\!30$ bits to $B_{tot} \!=\!3$ bits.

This observation can be explained as the following.  For small $P_{tot}$, the communication channels cannot support reliable transmission of large number of bits. Hence, the algorithms allocate few bits to sensors and increasing $B_{tot}$ does not improve the performance.
%
%
Another observation is that for $B_{tot}\!=\!30$ bits (plentiful bandwidth) and large $P_{tot}$, performance of all algorithms reaches the clairvoyant benchmark $d_0$, whereas for $B_{tot}\!=\!3$ bits (scarce bandwidth) and large $P_{tot}$ there is a persistent gap with $d_0$ for each algorithm, due to quantization errors, and  ``$a$-coupled'' and ``$b$-decoupled''  algorithms perform the best and the worst, respectively. 
%

Fig. \ref{MSE-simulated-vs-P_tot} depicts the Mont Carlo simulated MSE when  ``$a$-coupled'',   ``$b$-coupled'', ``$a$-decoupled'' and ``$b$-decoupled'' algorithms are implemented for power and rate allocation vs. $P_{tot}$ for $B_{tot} \!=\! 30, 3$ bits. Similar observations to those of Fig. \ref{MSE-upperbound-comparison} are made, with a few differences:
%
(i) ``$b$-coupled'' outperforms ``$a$-decoupled''  algorithm in very low $P_{tot}$, (ii) the simulated MSE obtained by ``$b$-coupled'', ``$a$-decoupled'', and  ``$b$-decoupled'' algorithms approaches the same value for $B_{tot} \!=\!3$ bits and large $P_{tot}$. The reason perhaps is that the discretized quantization rates are the same for these algorithms and according to \eqref{HP-power-allocation} for large $P_{tot}$, $P_k$'s of these algorithms become identical.

To better see the behavior of the upper bounds with respect to the simulated MSE,  Figs. \ref{MSE-and-upperbounds-vs-P_tot-A} and \ref{MSE-and-upperbounds-vs-P_tot-B} plot the simulated MSE and upper bounds vs. $P_{tot}$, for all the proposed algorithms (note that Figs. \ref{MSE-and-upperbounds-vs-P_tot-A} and \ref{MSE-and-upperbounds-vs-P_tot-B} have the same curves as  Figs. \ref{MSE-upperbound-comparison} and \ref{MSE-simulated-vs-P_tot}). 
We observe that when $P_{tot}$ and $B_{tot}$ are not too low, the upper bounds are good approximations of the simulated MSE.


Fig. \ref{MSE-upperbounds-vs-B_tot} depicts $\mathcal D_a$ when implementing ``$a$-coupled'' and ``$a$-decoupled''  and $\mathcal D_b$ when implementing ``$b$-coupled'' and ``$b$-decoupled'' algorithms,   vs. $B_{tot}$, for $P_{tot}\!=\! 30, 16$ dB. \\
Fig. \ref{MSE-simulated-vs-B_tot} depicts the simulated MSE when  ``$a$-coupled'',   ``$b$-coupled'', ``$a$-decoupled'' and ``$b$-decoupled'' algorithms are implemented for power and rate allocation vs. $B_{tot}$ for $P_{tot} \!=\! 30, 16$ dB.
%
%
We make similar observations and conclusions to those we made for Fig. \ref{MSE-simulated-vs-P_tot}.  For $P_{tot} \!=\!30$ dB and large $B_{tot}$, performance of all algorithms reaches the clairvoyant benchmark $d_0$. On the other hand, for $P_{tot} \!=\!16$ dB and large $B_{tot}$, there is a persistent gap with $d_0$ for each algorithm, due to communication channel errors.
When comparing Figs. \ref{MSE-upperbounds-vs-B_tot} and \ref{MSE-simulated-vs-B_tot}, we note that the behavior of the bounds $\mathcal D_a$ and $\mathcal D_b$ vs. $B_{tot}$ is very similar to that of the simulated MSE.  


\section{Conclusions}\label{conclusions}
\vspace{-0.1cm}
We considered distributed estimation of a Gaussian vector with a known covariance matrix and linear observation model, 
where the FC is tasked with reconstruction of the unknowns, using a linear estimator.  Sensors employ uniform multi-bit quantizers and BPSK modulation,  and communicate with the FC over power- and bandwidth-constrained channels. We derived two closed-form upper bounds on the MSE, in terms of the optimization parameters (i.e., transmit power and quantization rate per sensor). Each bound consists of two terms, where the first term is the MSE due to observation noises and quantization errors, and the second term is the MSE due to communication channel errors. We proposed ``coupled'' and ``decoupled'' resource allocation schemes that minimize these bounds. The ``coupled'' schemes  utilize the iterative modified ellipsoid method to conduct $K$-dimensional search and find the quantization rate vector,  whereas the ``decoupled'' ones rely on one-dimensional search to find the quantization rates. Our simulations show that when $P_{tot}$ and $B_{tot}$  are not too scarce, the bounds are good approximations of the actual MSE. Through simulations, we verified the effectiveness of the proposed schemes and confirmed that their performance approaches the clairvoyant centralized estimation for large $P_{tot}$ and $B_{tot}$  ($P_{tot}\!\approx\!25$ dB, $B_{tot} \!\approx\!30$ bits). Our results indicate that resource allocation is affected by sensors' observation qualities, channel gains, as well as $P_{tot}$ and $B_{tot}$, e.g., two WSNs with identical conditions and $P_{tot}$ ($B_{tot}$) and different $B_{tot}$ ($P_{tot}$) require two different power (rate) allocation. Also, more number of bits and transmit power are allotted to sensors with better observation qualities. 
\vspace{-0.4cm}
\section{Appendix}
\vspace{-0.1cm}
\subsection{Finding Upper Bound on $\mathbb{E}\{(\hat{m}_k-m_k)^2\}$}\label{upperboudn-m_prim-m-appnedix}
\vspace{-0.1cm}
Suppose the bit sequence representations of $m_k$ and $\hat{m}_k$ are, respectively,  $(b_{k,1}, ..., b_{k,L_k})$ and  $(\hat{b}_{k,1}, ..., \hat{b}_{k,L_k})$, i.e.,
$m_k\!=\!\Delta_k (0.5-2^{L_k-1}+\sum_{j=1}^{L_k} b_{k,j}2^{L_k-j})$ and
$\hat{m}_k\!=\!\Delta_k (0.5-2^{L_k-1}+\sum_{j=1}^{L_k} \hat{b}_{k,j} 2^{L_k-j})$. Therefore:
\vspace{-0.2cm}
\begin{eqnarray*}
\mathbb{E}\{ (\hat{m}_k-m_k)^2\} & = &\Delta_k^2 \mathbb{E}\{(\sum_{j=1}^{L_k} 2^{L_k-j}(b_{k,j}-\hat{b}_{k,j}))^2\}\nonumber \\
 & \overset{(a)}{\leq} & \Delta_k^2 L_k  (4^{L_k}) \sum_{j=1}^{L_k} 4^{-j} \underbrace{\mathbb{E}\{ (b_{k,j}-\hat{b}_{k,j})^2\}}_{=p_e}\nonumber\\
&\overset{(b)}{=}  &p_e \Delta_k^2 L_k (4^{L_k})\frac{1-(1/4)^{L_k}}{3}\\ 
&\overset{(c)}{<}  &\frac{ 4 p_e L_k \tau_k^2}{3}
 \overset{(d)}{\leq}  \mbox{exp}(-\frac{\gamma_k P_k}{L_k})\frac{ 4 L_k \tau_k^2}{3}\nonumber
\end{eqnarray*}
where $(a)$ comes from Cauchy's inequality $(\sum_j \alpha_j \beta_j)^2 \! \leq \! (\sum_j \alpha_j^2)(\sum_j \beta_j^2)$ for arbitrary $\alpha_j, \beta_j$'s, and the fact that 
$(b_{k,j}-\hat{b}_{k,j})^2$ is a Bernoulli random variable with success probability $p_e$,
%
$(b)$ is due to sum of a geometric series,
$(c)$  is found using the definition of $\Delta_k$ and $1-(1/4)^{L_k} \! < \! 1$,
and $(d)$ is obtained since $p_e\!=\!\mathcal{Q}(\sqrt{\frac{2\gamma_kP_k}{L_k}})$ and $\mathcal Q(x)\leq \mbox{exp}(-\frac{x^2}{2})$. 
\vspace{-0.3cm}
%
%
\subsection{Properties of $\mathcal {D}_2^{upb}$ and $\mathcal {D}_2^{uupb}$ }\label{convexity-of-D^{upb}-wrs-P_k}
One can verify  the following:
\begin{eqnarray}
\frac {\partial \mathcal {D}_2^{upb}}{\partial P_k} \!&\!=\!&\!-\alpha_k \gamma_k \mbox{exp}(- \frac{\gamma_k P_k}{L_k}) \leq 0, ~\forall k \nonumber\\
\frac {\partial^2 \mathcal {D}_2^{upb}}{\partial P_k \partial P_l}\!&\!=\!&\! \nonumber
\begin{cases}
    0 ,& \text{if } ~~~k\neq l\\
    \frac{\alpha_k \gamma^2_k}{L_k} \mbox{exp}(- \frac{\gamma_k P_k}{L_k} ) \! \geq \!0,              & \text{if}  ~~~k=l 
\end{cases}
\end{eqnarray}
These imply that $\mathcal {D}_2^{upb}$ is a decreasing function of $P_k$'s and also a jointly convex function of $P_k$'s.
Similarly:
\begin{eqnarray}
\frac {\partial \mathcal {D}_2^{uupb}}{\partial P_k}\!\!&\!\!= \!\!&\!\! \tilde \lambda [\frac{\partial \boldsymbol{\mathcal M'}}{\partial P_k}]_{k,k}\!=\!
-\tilde \lambda(\frac {4\tau_k^2\gamma_k}{3})\mbox{exp}(-\frac{\gamma_k P_k}{L_k}) \! \leq \! 0, ~\forall k \nonumber\\
    \frac {\partial^2 \mathcal {D}_2^{uupb}}{\partial P_k \partial P_l}\!\!&\!\!= \!\!&\!\!\nonumber \begin{cases}
    0 ,& \text{if } ~~~k\neq l\\
    \tilde \lambda(\frac {4\tau_k^2\gamma^2_k}{3L_k}) \mbox{exp}(-\frac{\gamma_k P_k}{L_k}) \! \geq \!0,              & \text{if}  ~~~k=l 
\end{cases}
\end{eqnarray}
These imply that $\mathcal {D}_2^{uupb}$ is a decreasing function of $P_k$'s and also a jointly convex function of $P_k$'s.
\vspace{-0.3cm}
%
\subsection{Properties of  $\sum_{k=1}^{K} \delta_k \sigma^2_{\epsilon_k}$} \label{proof-convexity-D_1^{upb}}
One can verify the following:
\begin{eqnarray*}
\frac {\partial (\sum_{k=1}^{K} \delta_k \sigma^2_{\epsilon_k})}{\partial L_k}\!\!&\!\!= \!\!&\!\! -\frac{2 \ln 2 ~\delta_k \tau^2_k 2^{L_k}}{3(2^{L_k}-1)^3} \! \leq \! 0, ~\forall k \nonumber\\
\frac {\partial^2 (\sum_{k=1}^{K} \delta_k \sigma^2_{\epsilon_k})}{\partial L_k \partial L_l}\!\!&\!\!= \!\!&\!\! 
\begin{cases}
    0 ,& \text{if } ~~~k\neq l\\
    \frac{(\ln 2)^2 \delta_k \tau^2_k 2^{L_k+1} (1+2^{L_k+1})}{3(2^{L_k}-1)^4} \! \geq \!0,     & \text{if}  ~~~k=l 
\end{cases}
\end{eqnarray*}
These imply that $\sum_{k=1}^{K} \delta_k \sigma^2_{\epsilon_k}$ is a decreasing function of $L_k$'s and also a jointly convex function of $L_k$'s.
\vspace{-0.3cm}
%
\subsection{More Properties on $\mathcal D_1$ and $\mathcal D_2^{upb}$ }\label{properties-on-distortion-elements}
$\bullet$ $\mathcal D_1$ is a decreasing function of $B_{tot}$:  after some mathematical manipulations, one can show
\begin{eqnarray} \label{derivative-D1-B_tot_appendix}
\frac{\partial{\mathcal{D}_1}}{\partial{B_{tot}}}=\tr(\boldsymbol G\frac{\partial \boldsymbol Q}{\partial B_{tot}} \boldsymbol G^T)=\sum_{k=1}^{K}[\frac{\partial \boldsymbol Q}{\partial B_{tot}}]_{k,k} || {\boldsymbol g}_k||^2 \! \leq \! 0, \nonumber
\end{eqnarray}
where after substituting  \eqref{L_i-solution-final} into $\boldsymbol Q$
we find $[\frac{\partial \boldsymbol Q}{\partial B_{tot}}]_{k,k}\! =\!-\frac{2\ln 2 \tau_k^2 2^{L_k}}{3(2^{L_k}-1)^3 K} \! \leq \! 0 $.
As $B_{tot}  \! \rightarrow \! 0$, $[\boldsymbol Q]_{k,k} \! \rightarrow \! \infty$, i.e.,
all eigenvalues of $(\mathcal C_{\boldsymbol x }+\boldsymbol Q)$ go to infinity. Consequently, due to Weyl's inequality \cite{Rajendra} all eigenvalues of $(\mathcal C_{\boldsymbol x }+\boldsymbol Q)^{-1}$ go to
zero. Since $(\mathcal C_{\boldsymbol x}+\boldsymbol Q)^{-1}$ 
 is a diagonalizable matrix, this means $(\mathcal C_{\boldsymbol x}+\boldsymbol Q)^{-1}$ goes to an all-zero matrix and 
$\mathcal D_1 \! \rightarrow \! \tr(\mathcal C_{\bold \theta})$. 
On the other hand, as 
$B_{tot}  \! \rightarrow \! \infty$, $[\boldsymbol Q]_{k,k} \! \rightarrow \! 0$ and thus 
$\mathcal D_1 \! \rightarrow \! d_0 \! = \! \tr(\mathcal C_{\bold \theta}) \! - \!\tr(\mathcal C_{\boldsymbol x \bold \theta }^T \mathcal C_{\boldsymbol x}^{-1} \mathcal C_{\boldsymbol x \bold \theta })$. 
%

$\bullet$ $\mathcal D_2^{upb}$ is an increasing function of $B_{tot}$: after some mathematical manipulations, one can verify
\begin{eqnarray}\label{D2-upb-appendix-D}
\frac{\partial{\mathcal{D}_2^{upb}}}{\partial{B_{tot}}} =\tr ( \boldsymbol G  \frac{\partial \boldsymbol{\mathcal  {M'}}}{\partial B_{tot}}  \boldsymbol G^T )+ \tr (\boldsymbol G (\bold B+\bold B^T) \boldsymbol G^T),
\end{eqnarray}
where $\bold B\!=\!-(\frac{\partial \boldsymbol Q}{\partial B_{tot}})(\mathcal{C}_{\boldsymbol{x}}\!+\!\boldsymbol{Q})^{-1} \boldsymbol {\mathcal M'}$. Substituting  \eqref{L_i-solution-final} into $\boldsymbol {\mathcal M'}$ we find $[\frac{\partial \boldsymbol {\mathcal M'}}{\partial B_{tot}}]_{k,k} \! =  \!  \frac{\partial u_k }{\partial B_{tot}} \! =\! \frac{4 \tau_k^2}{3K}\mbox{exp}(\frac{- \gamma_k P_k }{ L_k})[1+ \frac{\gamma_k P_k}{L_k}] \! \geq \! 0 $. Hence, $\boldsymbol G  \frac{\partial \boldsymbol{\mathcal  {M'}}}{\partial B_{tot}}  \boldsymbol G^T  \! \succeq  \! 0 $ and the first term in (\ref{D2-upb-appendix-D}) in non-negative. 
Next we show that the second term in (\ref{D2-upb-appendix-D}) is also non-negative and hence 
$\frac{\partial{\mathcal{D}_2^{upb}}}{\partial{B_{tot}}} \! \geq \! 0$.
Since $\boldsymbol G^T \boldsymbol G$ and $\bold B \! +\! \bold B^T$ are symmetric matrices and $\boldsymbol G^T \boldsymbol G  \! \succeq  \! 0$, using inequality (2) of \cite{trace-inequality-Baksalary} we obtain:
 \begin{equation*}
\tr (\boldsymbol G (\bold B+\bold B^T) \boldsymbol G^T) \! = \! \tr(\boldsymbol G \boldsymbol G^T (\bold B+\bold B^T)) \geq  \lambda_{min} (\boldsymbol G \boldsymbol G^T) \tr(\bold B+\bold B^T)  
 \end{equation*} 
where $\lambda_{min} (\boldsymbol G \boldsymbol G^T) \! \geq \! 0$. Furthermore,
\begin{eqnarray*}\label{appendix-D-tr-lower-bound}
\tr(\bold B+\bold B^T) \!\! & \!\! = \!\! & \!\!  2 \tr(\bold B)
=2\tr(  \boldsymbol {\mathcal M'}(-\frac{\partial \boldsymbol Q}{\partial B_{tot}})   (\mathcal{C}_{\boldsymbol{x}}\!+\!\boldsymbol{Q})^{-1}) \nonumber\\
\!\! & \!\! \overset{(e)}{\geq}  \!\! & \!\!  2 \lambda_{min}(\boldsymbol {\mathcal M'}(-\frac{\partial \boldsymbol Q}{\partial B_{tot}}))\tr((\mathcal{C}_{\boldsymbol{x}}\!+\!\boldsymbol{Q})^{-1}) \! \overset{(f)}{\geq}  \! 0
\end{eqnarray*}
in which $(e),(f)$ above are found since $\boldsymbol {\mathcal M'}(-\frac{\partial \boldsymbol Q}{\partial B_{tot}})$ and
$(\mathcal{C}_{\boldsymbol{x}}\!+\!\boldsymbol{Q})^{-1}$ are symmetric and positive definite matrices. \\
As $B_{tot}  \! \rightarrow \! 0$, $[\boldsymbol {\mathcal M'}]_{k,k} \! \rightarrow \! 0$ and $\boldsymbol G, \boldsymbol G^T$ go to all-zero matrices and therefore $\mathcal D_2^{upb} \! \rightarrow \! 0$. On the other hand, as $B_{tot} \! \rightarrow \! \infty$, $[\boldsymbol {\mathcal M'}]_{k,k} \! \rightarrow \! \infty$ and $\boldsymbol G \! \rightarrow \! \mathcal{C}_{{\boldsymbol{x}}{\boldsymbol{\theta}}}^T \mathcal{C}_{\boldsymbol{x}}^{-1}$ and therefore $\mathcal D_2^{upb} \! \rightarrow \! \infty$.
%
%


\begin{figure}[h!]
 \centering
\subcaptionbox{$B_{tot}=30$ (bits)}{\vspace{-.25 cm}\includegraphics[width=3.4in]{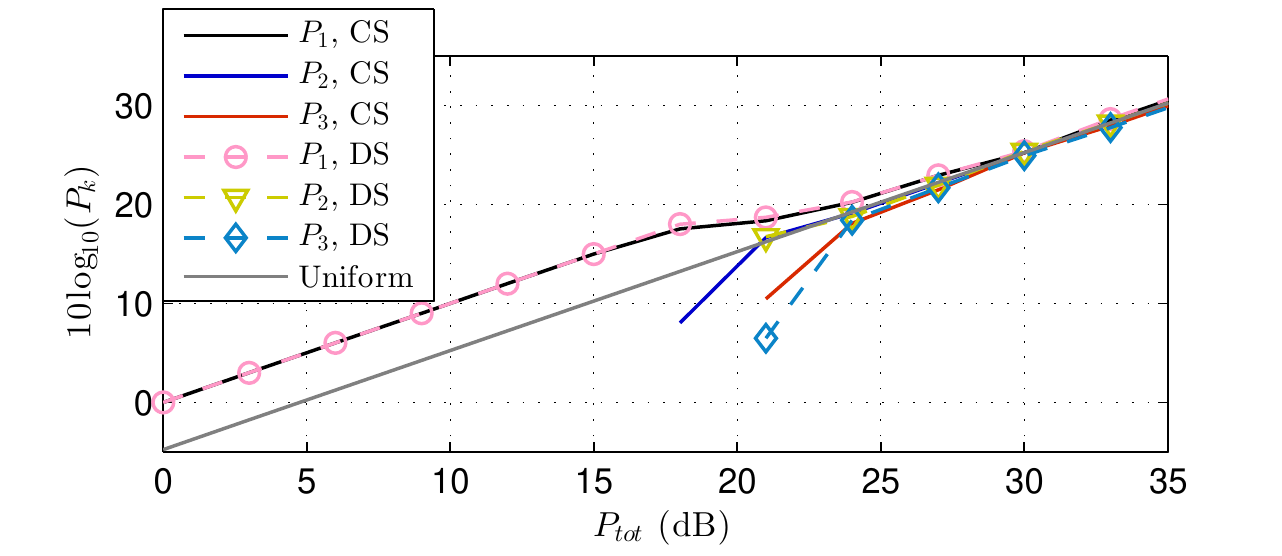}}\\
     \vspace{-.07cm}
     \subcaptionbox{$B_{tot}=3$ (bits)}{\vspace{-.25 cm}\includegraphics[width=3.4in]{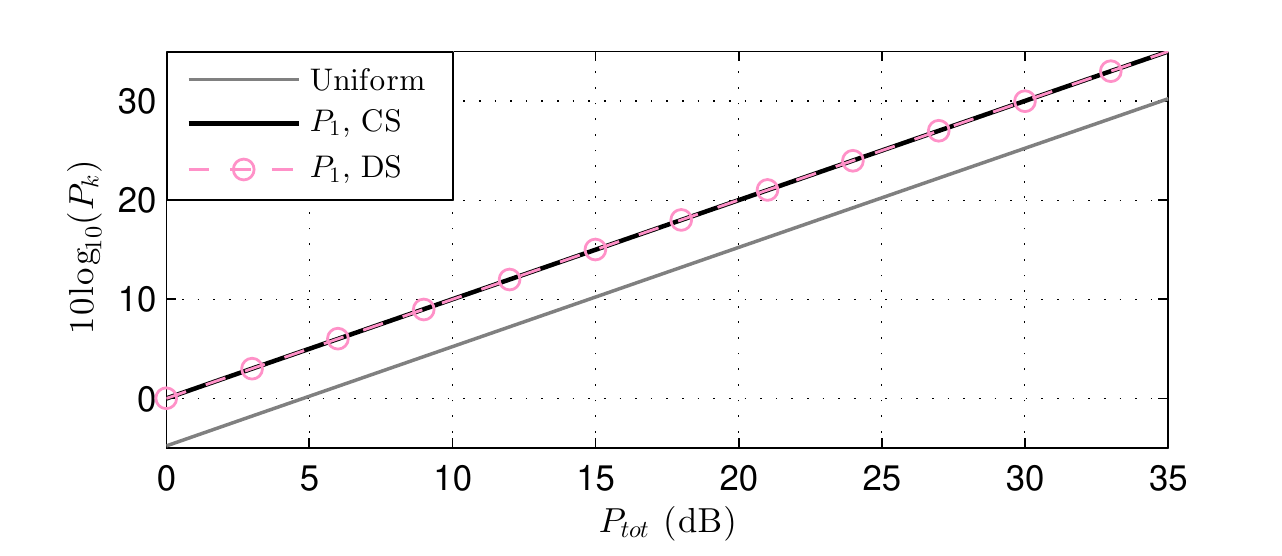}}
     \vspace{-.1cm}     
     \caption{``$a$-coupled'' algorithm $\{ 10 \log_{10} (P_k)\}_{k=1}^3$ vs. $P_{tot}$}   
      \label{power-vs-power-coupled-1}    
 \end{figure}

\begin{figure}[h!]
 \centering
     \subcaptionbox{$B_{tot}=30$ (bits)}{\vspace{-.25 cm}\includegraphics[width=3.4in]{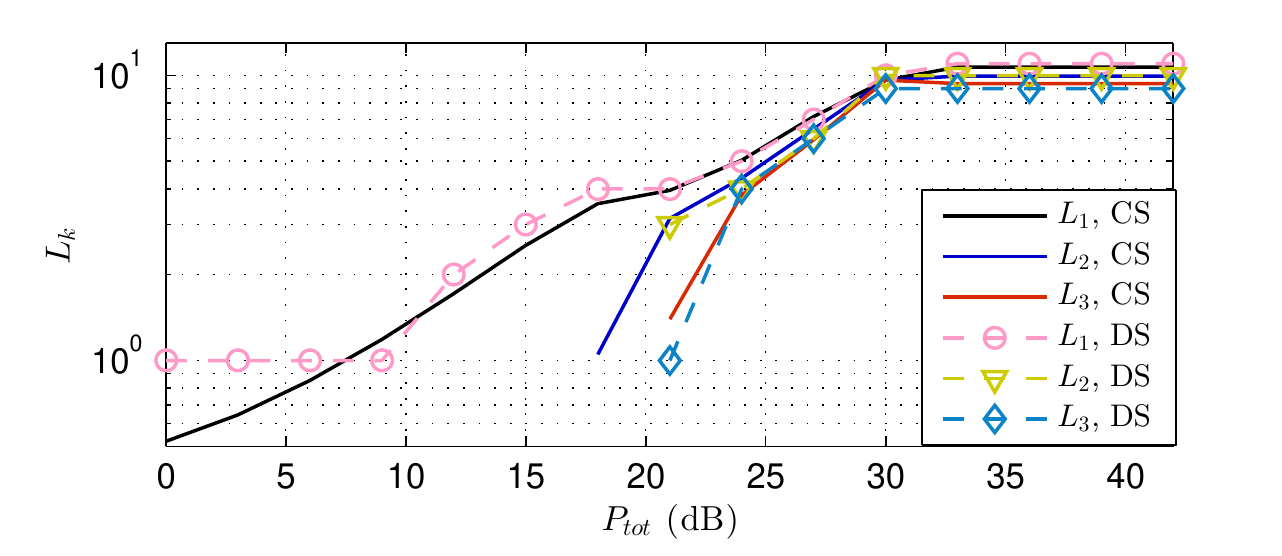}}\\
     \vspace{-.07cm}
     \subcaptionbox{$B_{tot}=3$ (bits)}{\vspace{-.25 cm}\includegraphics[width=3.4in]{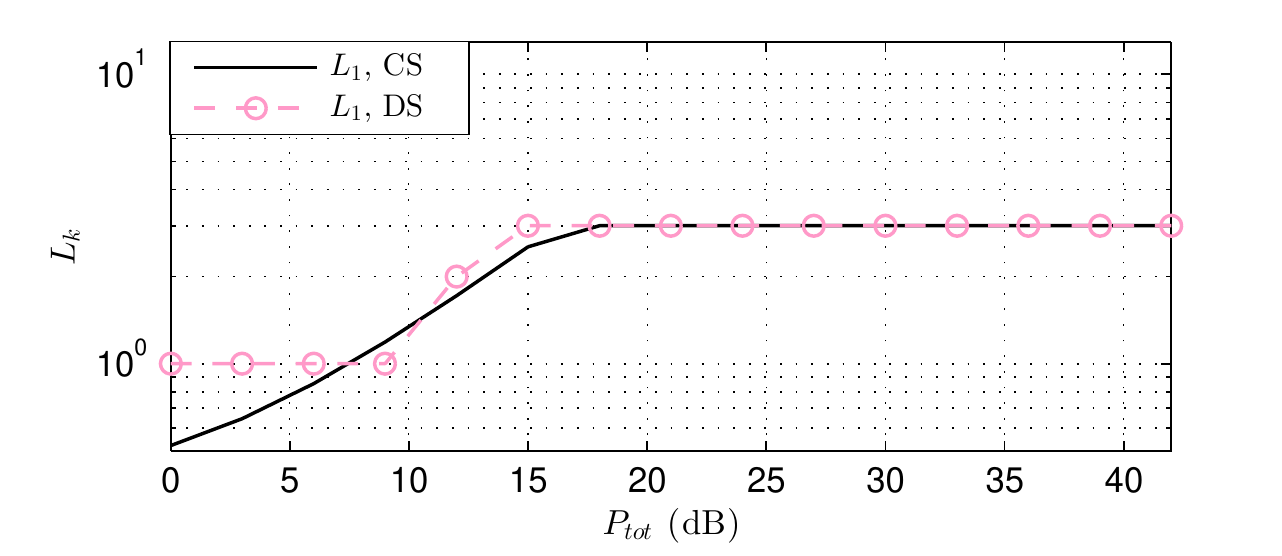}}
     \vspace{-.1cm}
     \caption{``$a$-coupled'' algorithm $\{L_k\}_{k=1}^3$ vs. $P_{tot}$}   
      \label{rate-vs-power-Coupled-1}   
 \end{figure}
\begin{figure}[h!]
 \centering
     
     \subcaptionbox{$B_{tot}=30$ (bits)}{\vspace{-.25 cm}\includegraphics[width=3.4in]{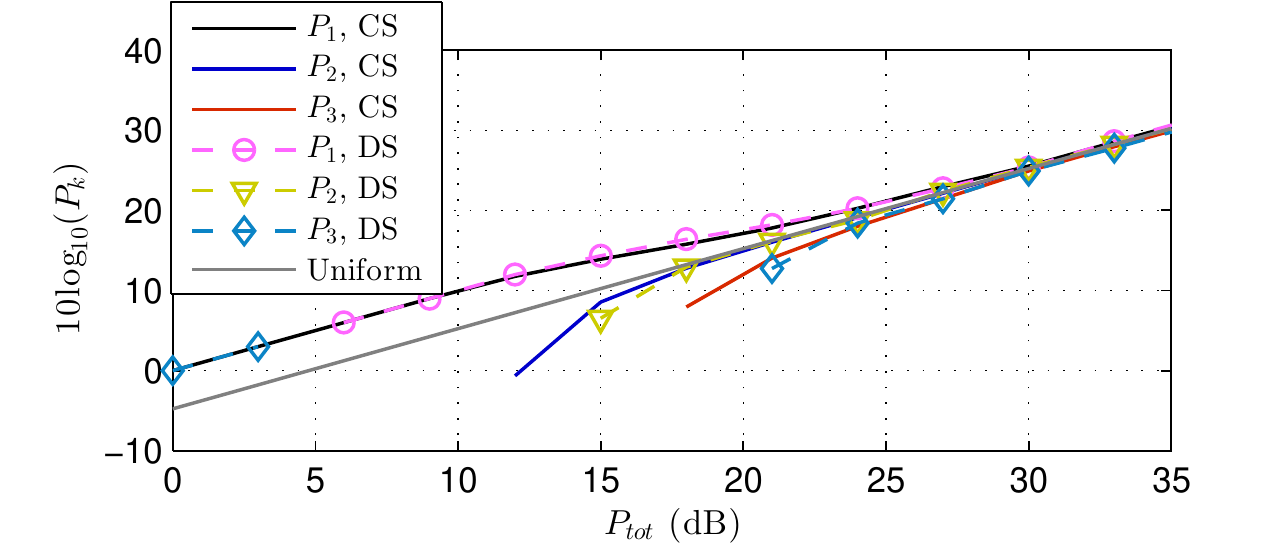}}\\
          \vspace{-.07cm}
          \subcaptionbox{$B_{tot}=3$ (bits)}{\vspace{-.25 cm}\includegraphics[width=3.4in]{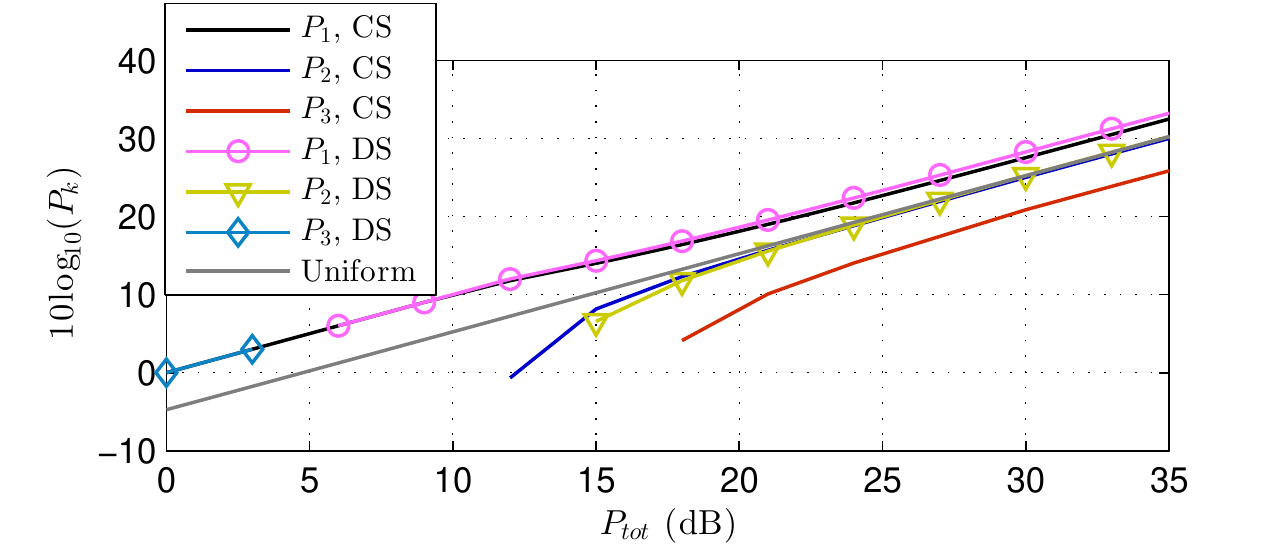}}
          \vspace{-.1cm}
     
     \caption{``$a$-decoupled'' algorithm $\{ 10 \log_{10} (P_k)\}_{k=1}^3$ vs. $P_{tot}$}   
      \label{power-vs-power-decoupled-1}    
 \end{figure}
\begin{figure}[h!]
 \centering

     \subcaptionbox{$B_{tot}=30$ (bits)}{\vspace{-.25 cm}\includegraphics[width=3.4in]{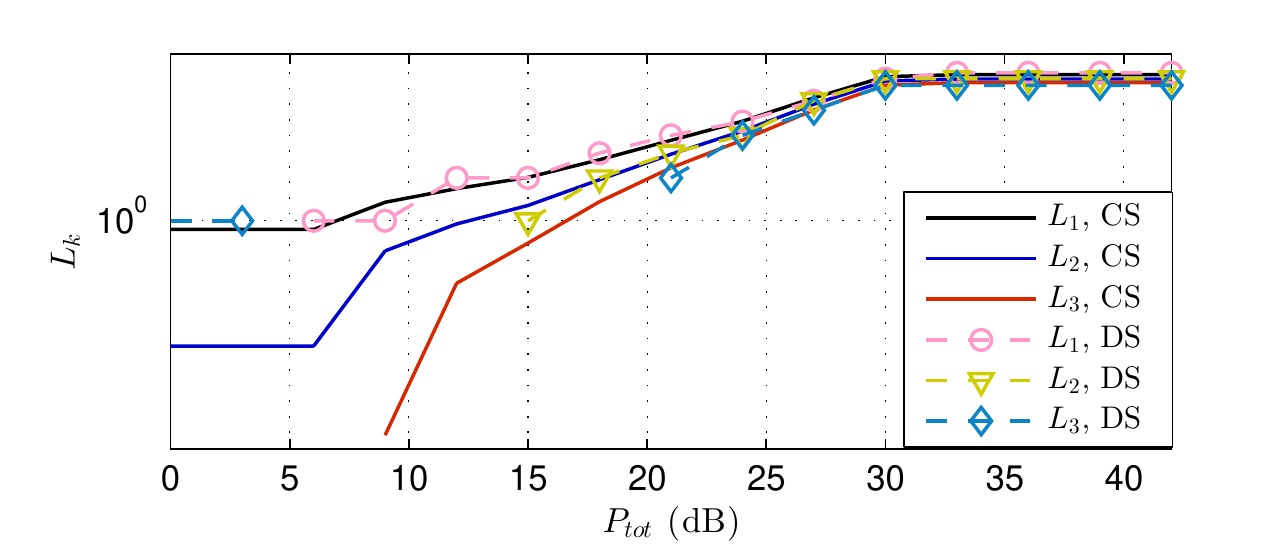}}\\
               \vspace{-.07cm}
               \subcaptionbox{$B_{tot}=3$ (bits)}{\vspace{-.25 cm}\includegraphics[width=3.4in]{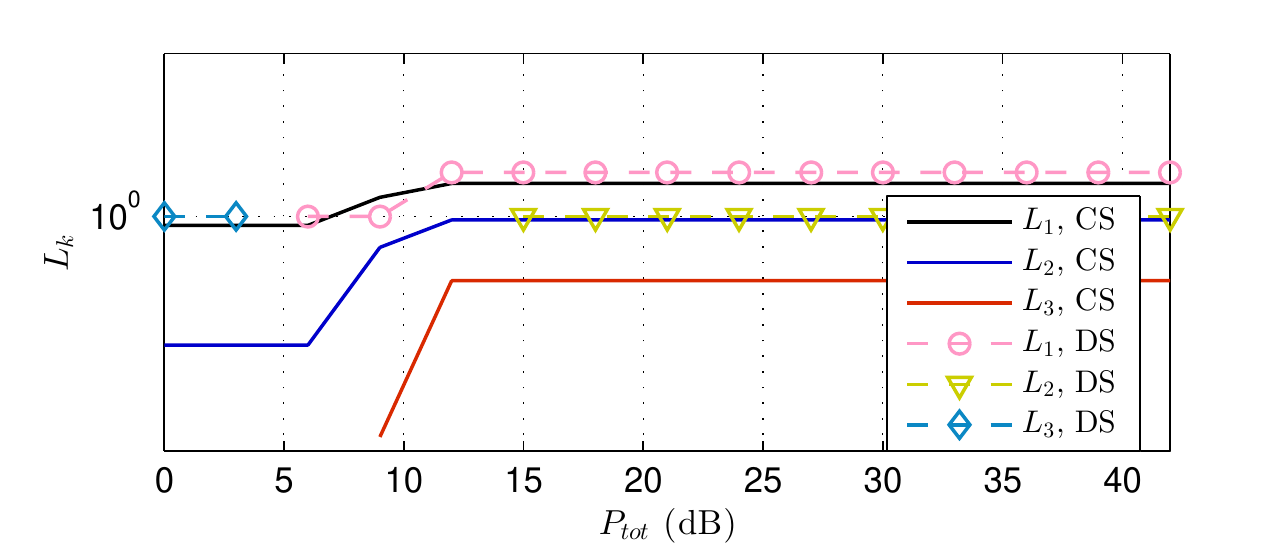}}
               \vspace{-.1cm}
          
     \caption{``$a$-decoupled'' algorithm $\{L_k\}_{k=1}^K$ vs. $P_{tot}$}   
      \label{rate-allocation-A-decoupled-1}     
 \end{figure}
\begin{figure}[h!]
             \includegraphics[width=3.5in]{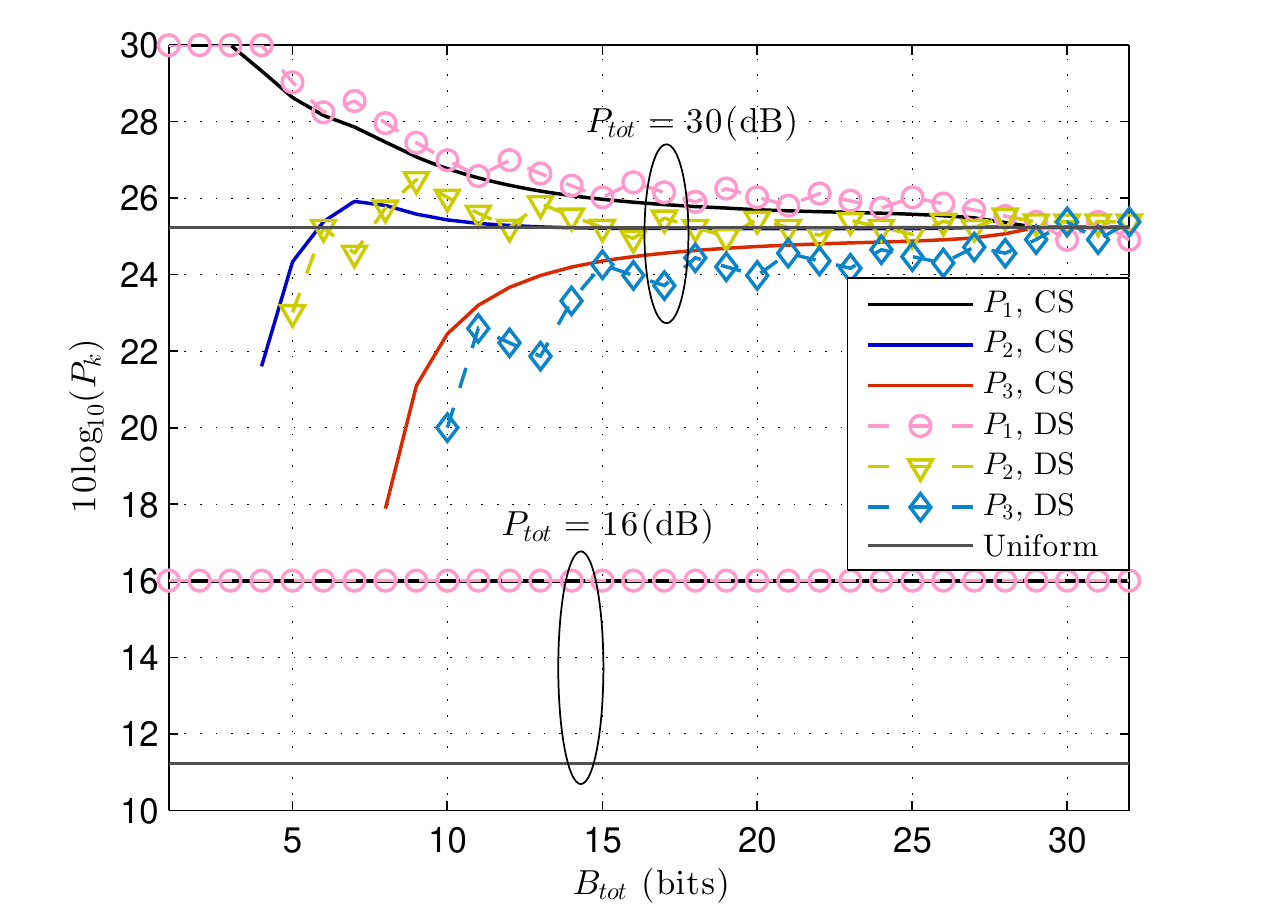}
             \vspace{-0.5cm}
             \caption{``$a$-coupled'' algorithm $\{ 10 \log_{10} (P_k)\}_{k=1}^3$ vs. $B_{tot}$}
             \label{power-vs-BW-coupled-1}
             \vspace{-.2cm}
             \end{figure}
\begin{figure}[h!]
             \includegraphics[width=3.5in]{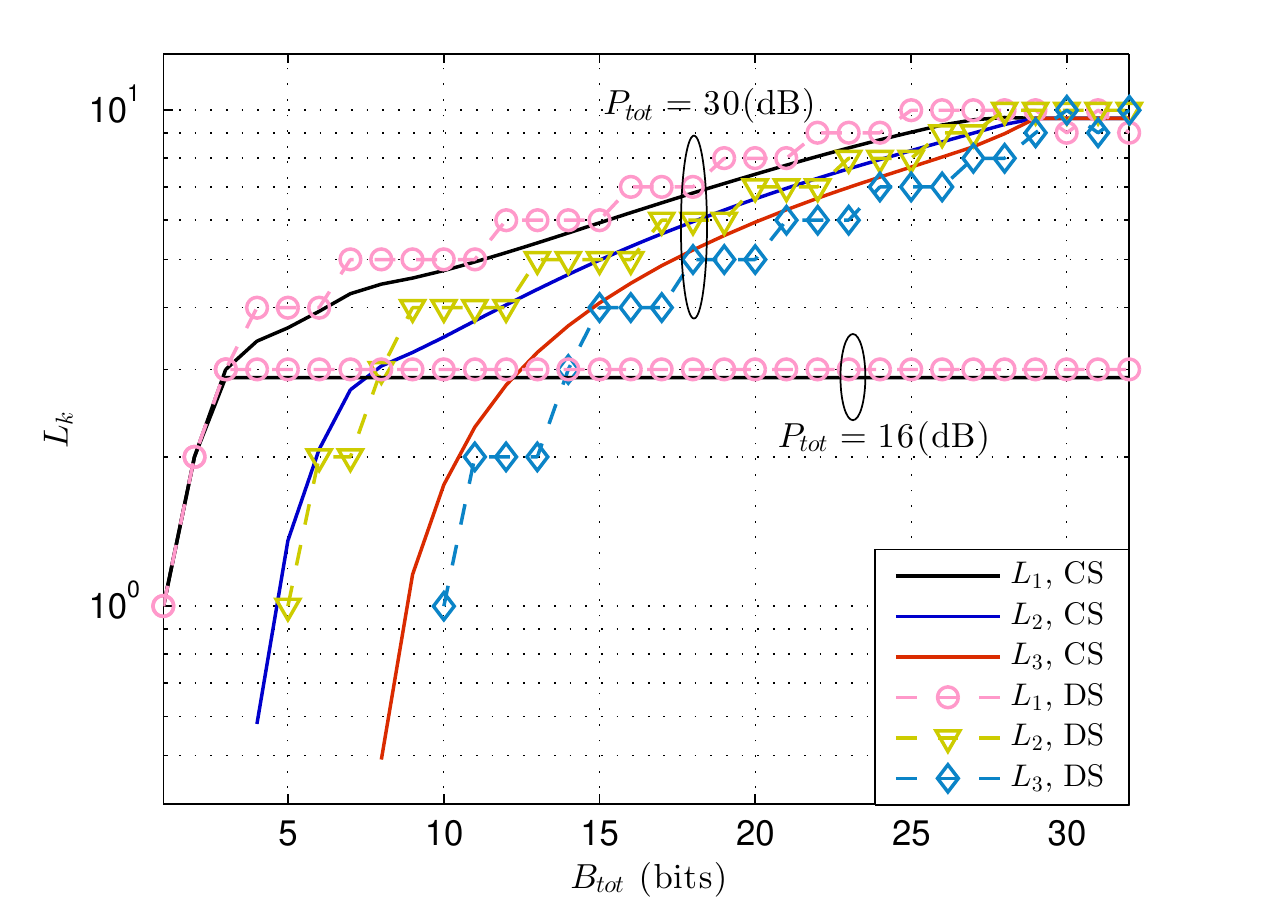}
             \vspace{-0.5cm}
             \caption{``$a$-coupled'' algorithm $\{L_k\}_{k=1}^K$ vs. $B_{tot}$}
             \label{rate-vs_BW-Coupled-1}
             \vspace{-.2cm}
\end{figure}
\begin{figure}[h!]
             \includegraphics[width=3.5in]{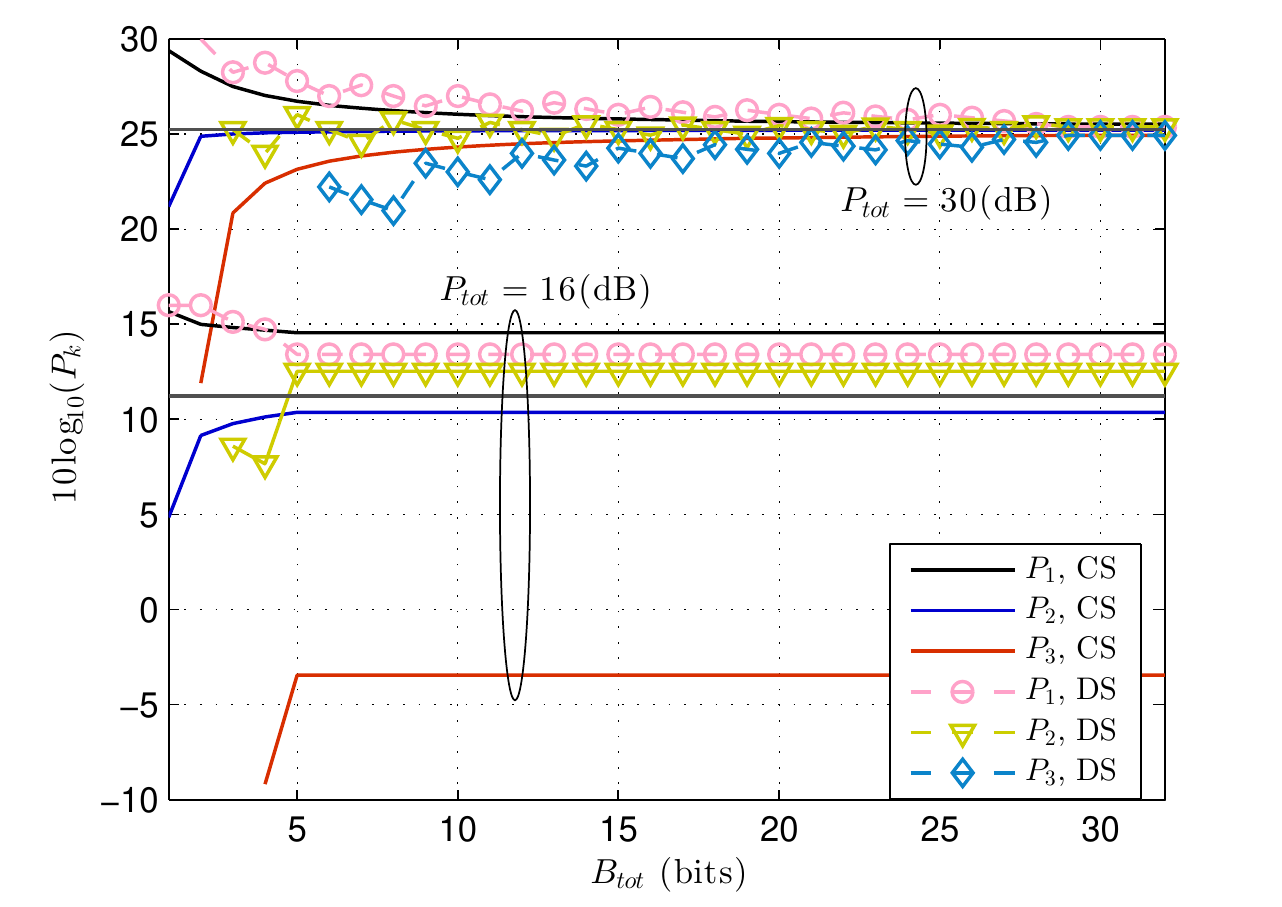}
             \vspace{-0.5cm}
             \caption{``$a$-decoupled'' algorithm $\{ 10 \log_{10} (P_k)\}_{k=1}^3$ vs. $B_{tot}$}
             \label{power-vs_BW-decoupled-1}
             \vspace{-.2cm}
\end{figure}
%

\begin{figure}[h!]
             \includegraphics[width=3.5in]{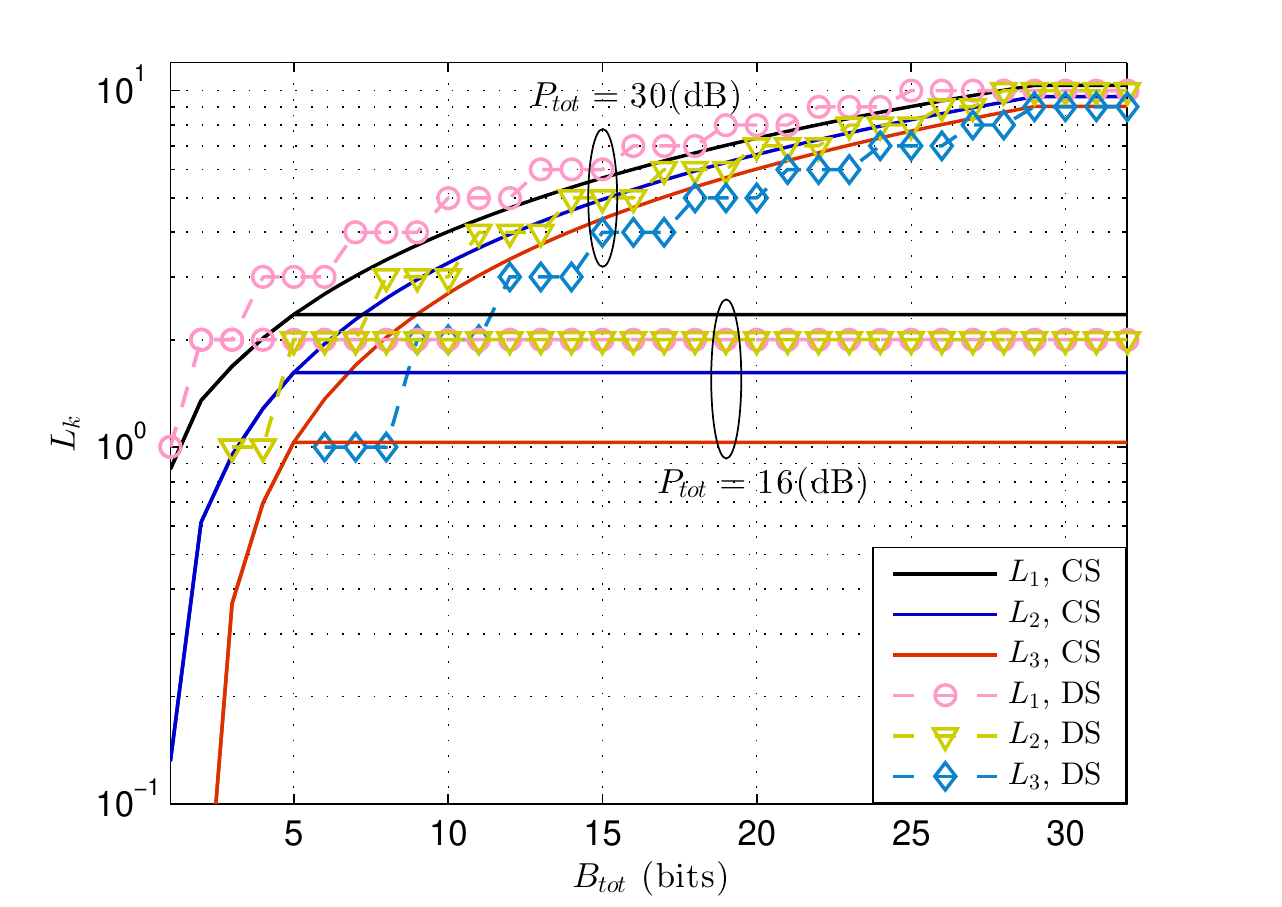}
             \vspace{-0.5cm}
             \caption{``$a$-decoupled'' algorithm $\{L_k\}_{k=1}^K$ vs. $B_{tot}$}
             \label{rate-vs_BW-decoupled-1}
             \vspace{-.2cm}
\end{figure}

\begin{figure}[h!]
 \centering

     \subcaptionbox{$B_{tot}=30$ (bit)}{\vspace{-.25 cm}\includegraphics[width=3.4in]{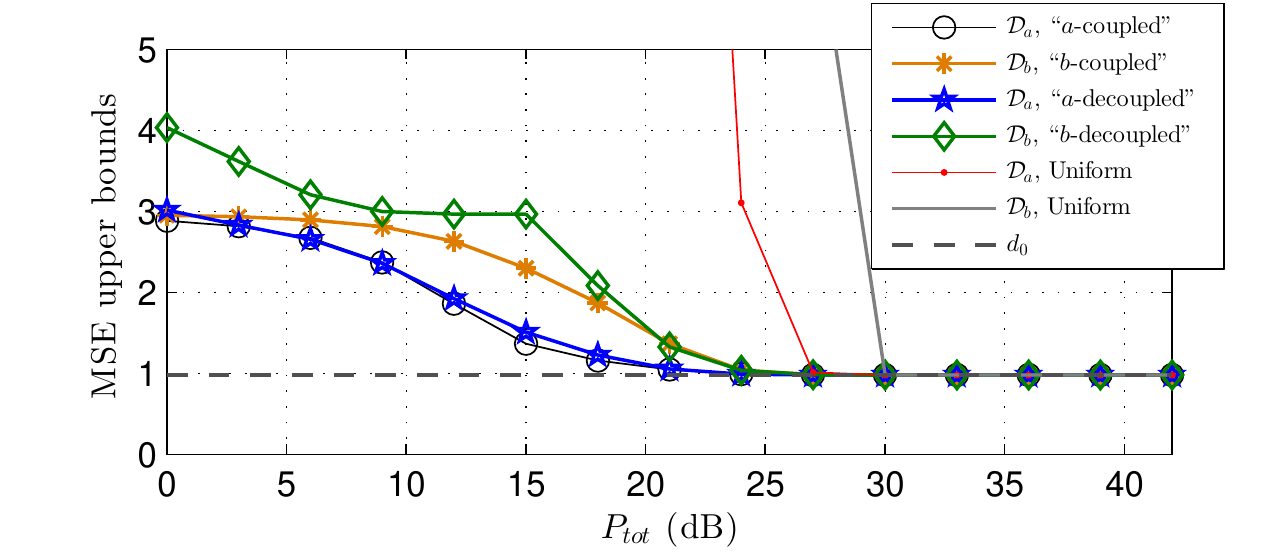}}\\
                    \vspace{-.07cm}
                    \subcaptionbox{$B_{tot}=3$ (bit)}{\vspace{-.25 cm}\includegraphics[width=3.4in]{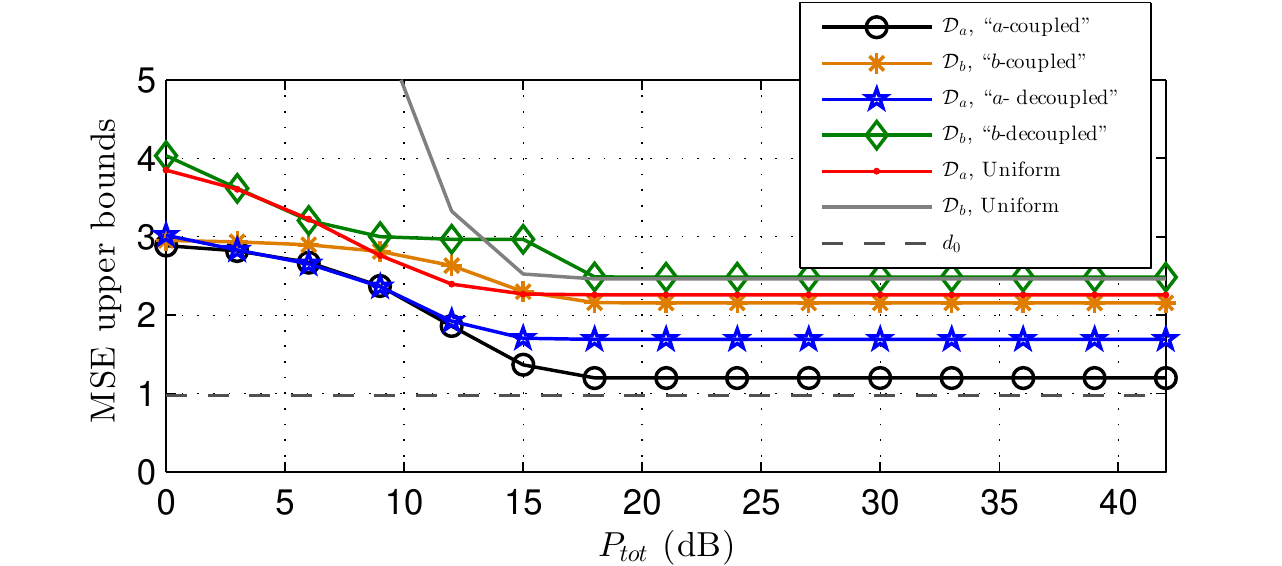}}
                    \vspace{-.1cm}
          
     \caption{$\mathcal D_a$ and $\mathcal D_b$ vs. $P_{tot}$ for all algorithms}   
      \label{MSE-upperbound-comparison} 
    
 \end{figure}

 \begin{figure}[h!]
  \centering

      \subcaptionbox{$B_{tot}=30$ (bits)}{\vspace{-.25 cm}\includegraphics[width=3.4in]{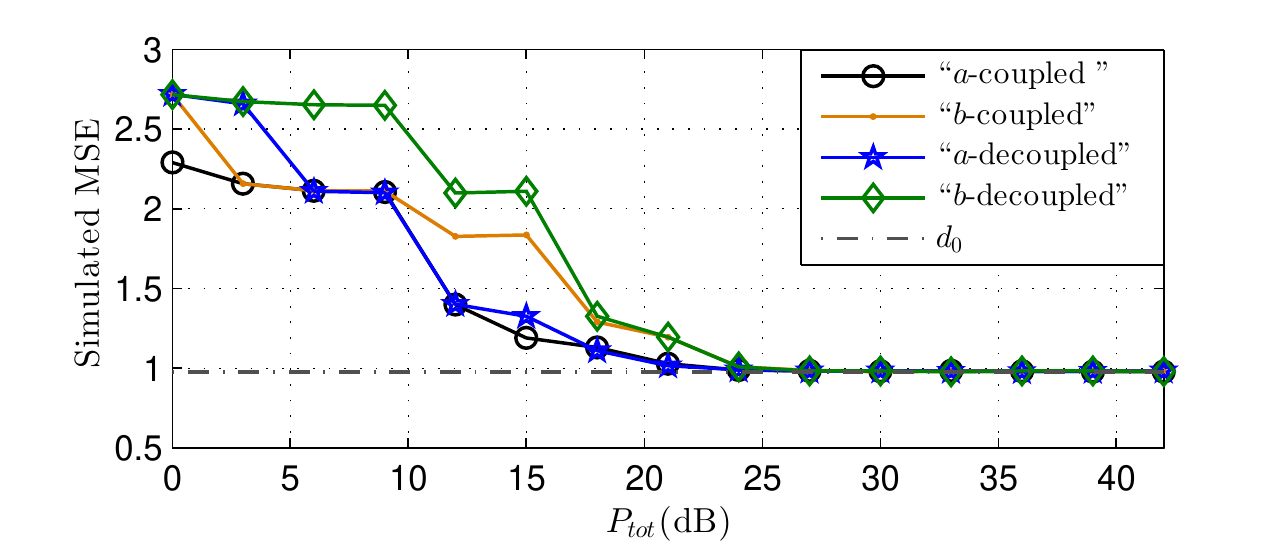}}\\
                          \vspace{-.07cm}
                          \subcaptionbox{$B_{tot}=3$ (bits)}{\vspace{-.25 cm}\includegraphics[width=3.4in]{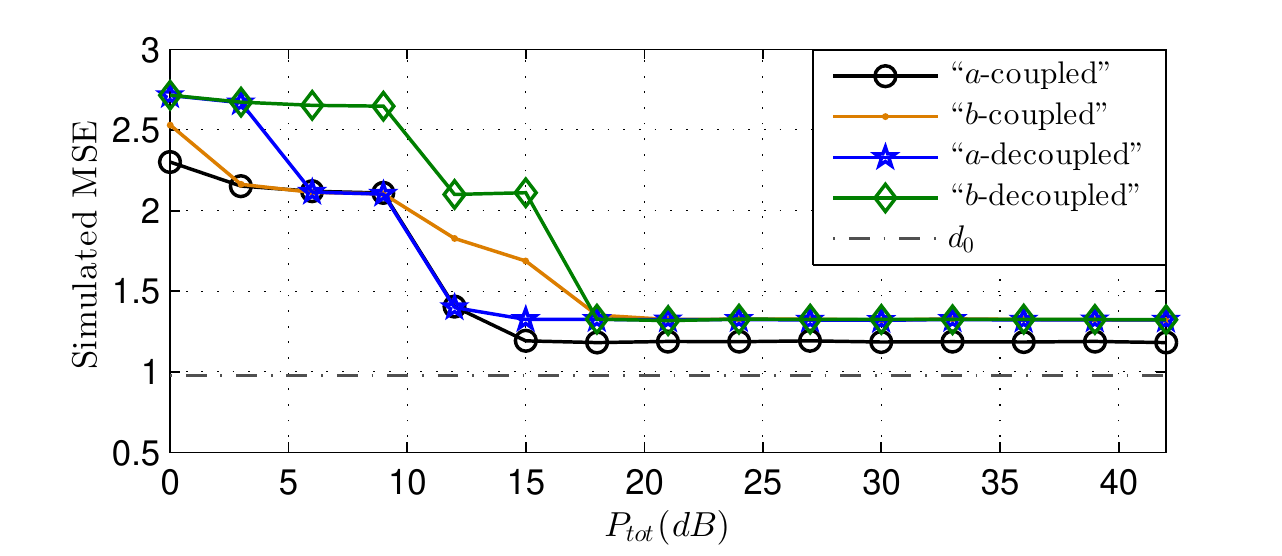}}
                          \vspace{-.1cm}
      \caption{Simulated MSE vs. $P_{tot}$ for all algorithms}   
       \label{MSE-simulated-vs-P_tot} 
     
  \end{figure}

 \begin{figure}[h!]
              \includegraphics[width=3.5in]{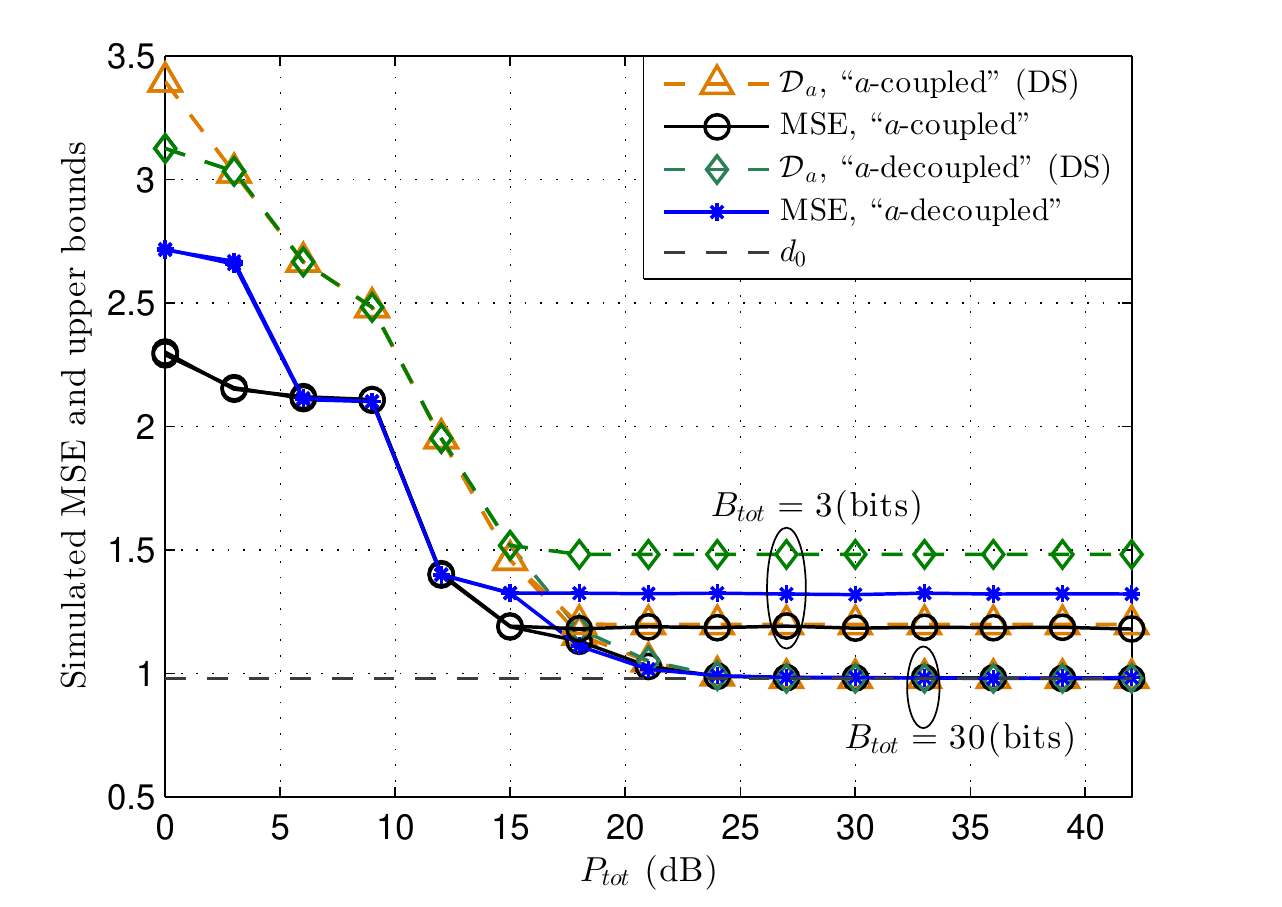}
              \vspace{-0.5cm}
              \caption{Simulated MSE and $\mathcal D_a$ vs. $P_{tot}$  for ``$a$-coupled'' and ``$a$-decoupled'' algorithms}
              \label{MSE-and-upperbounds-vs-P_tot-A}
              \vspace{-.2cm}
 \end{figure}
 
 \begin{figure}[h!]
               \includegraphics[width=3.5in]{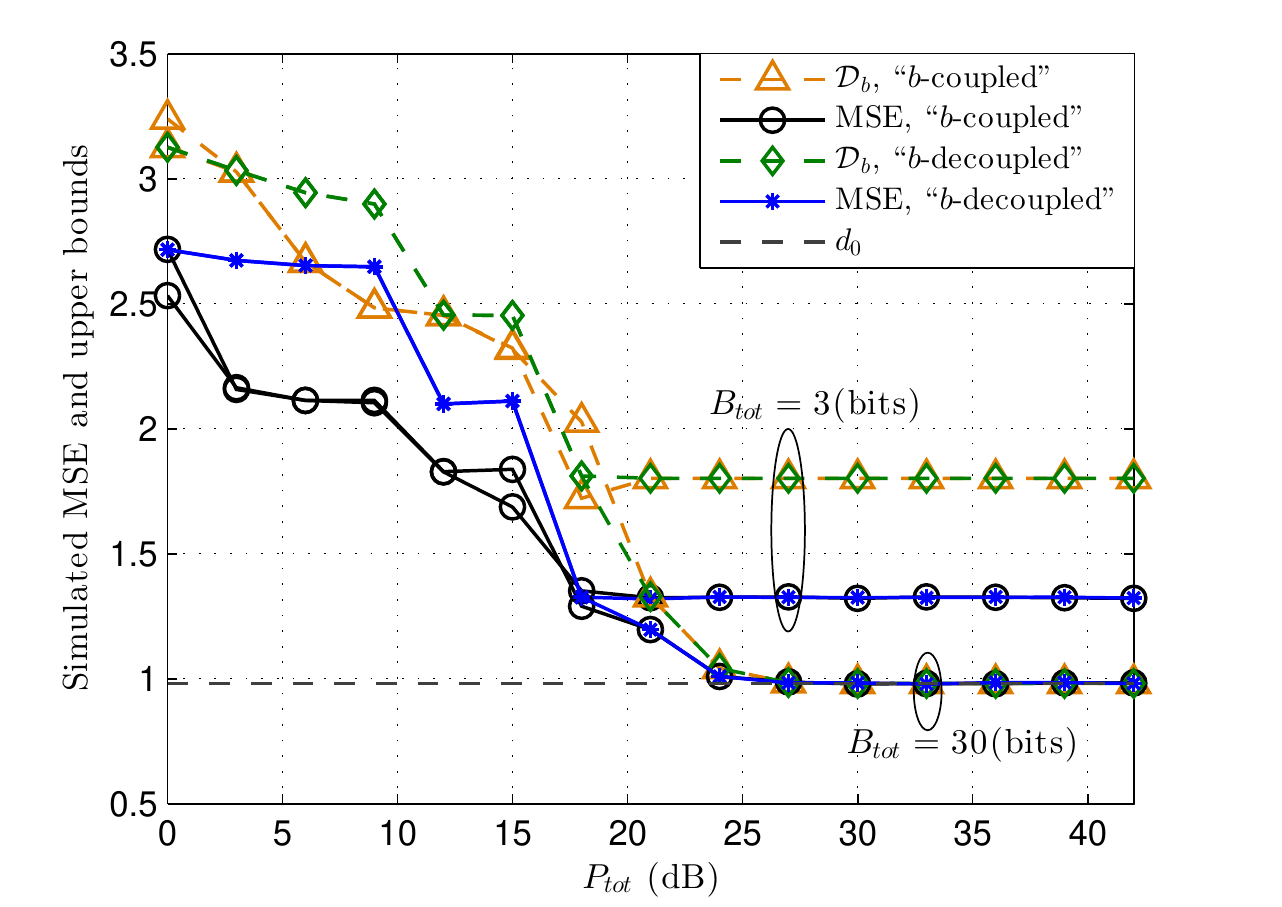}
               \vspace{-0.5cm}
               \caption{Simulated MSE and $\mathcal D_b$ vs. $P_{tot}$, for ``$b$-coupled'' and ``$b$-decoupled'' algorithms}
               \label{MSE-and-upperbounds-vs-P_tot-B}
               \vspace{-.2cm}
  \end{figure}

 \begin{figure}[h!]
    \centering

        \subcaptionbox{$P_{tot}=30$ (dB)}{\vspace{-.25 cm}\includegraphics[width=3.4in]{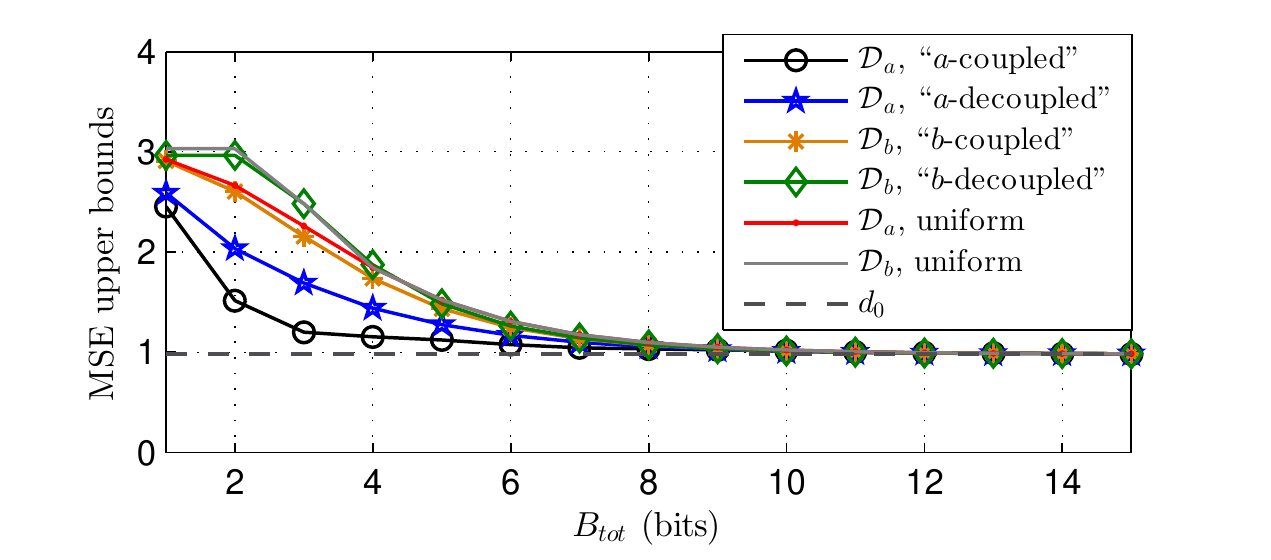}}\\
                                  \vspace{-.07cm}
                                  \subcaptionbox{$P_{tot}=16$ (dB)}{\vspace{-.25 cm}\includegraphics[width=3.4in]{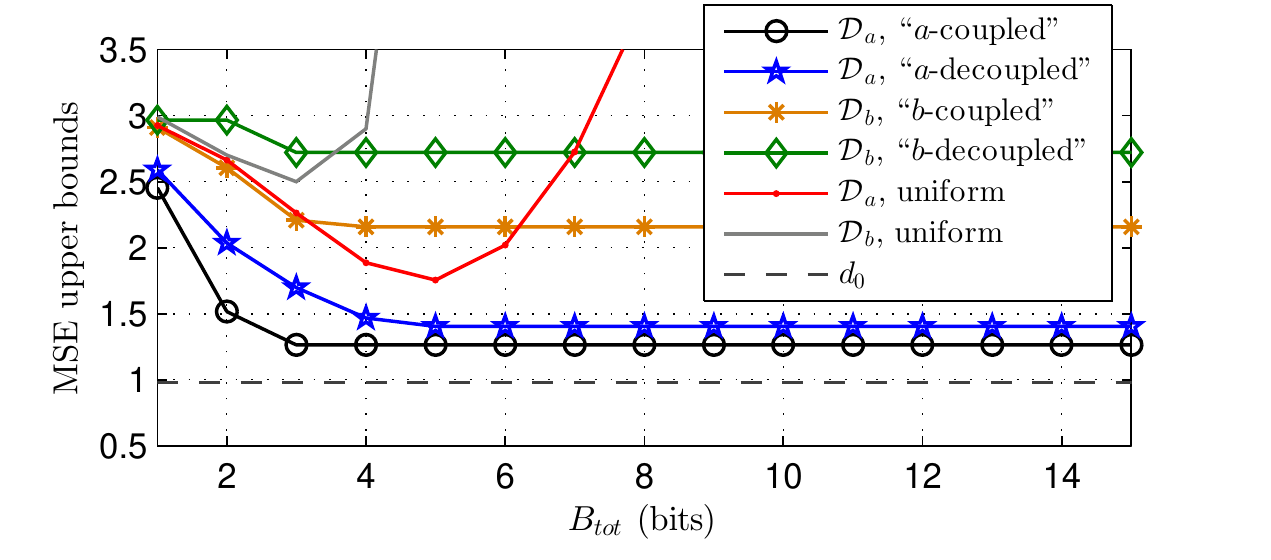}}
                                  \vspace{-.1cm}

        \caption{$\mathcal D_a$ and $\mathcal D_b$ vs. $B_{tot}$ for all algorithms}   
         \label{MSE-upperbounds-vs-B_tot} 
       
    \end{figure}
  
 \begin{figure}[h!]
    \centering

        \subcaptionbox{$P_{tot}=30$ (dB)}{\vspace{-.25 cm}\includegraphics[width=3.4in]{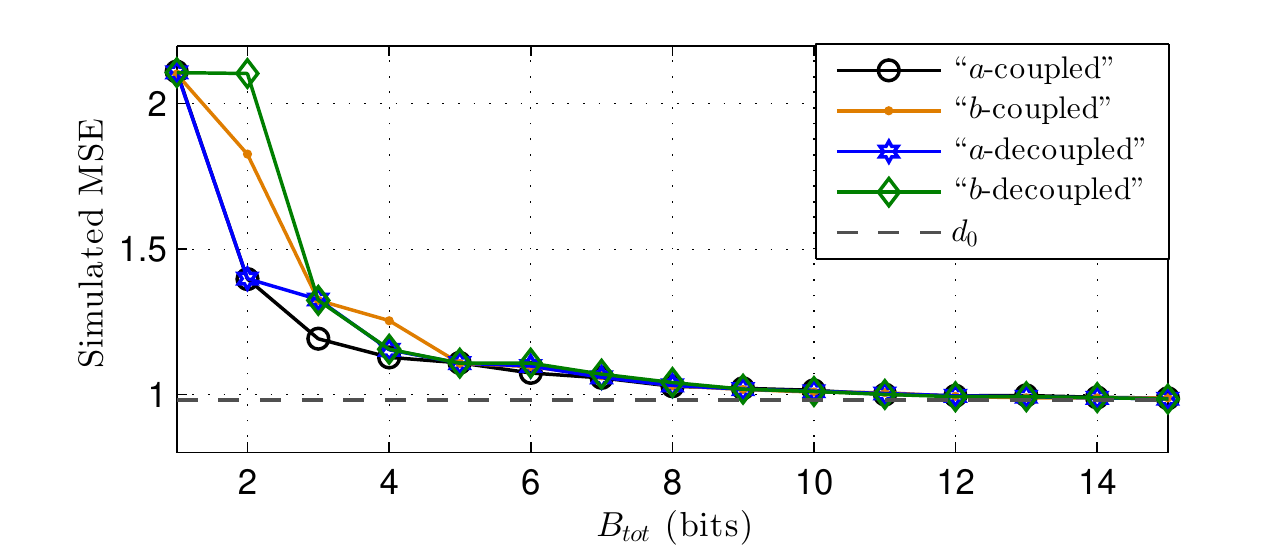}}\\
                                  \vspace{-.07cm}
                                  \subcaptionbox{$P_{tot}=16$ (dB)}{\vspace{-.25 cm}\includegraphics[width=3.4in]{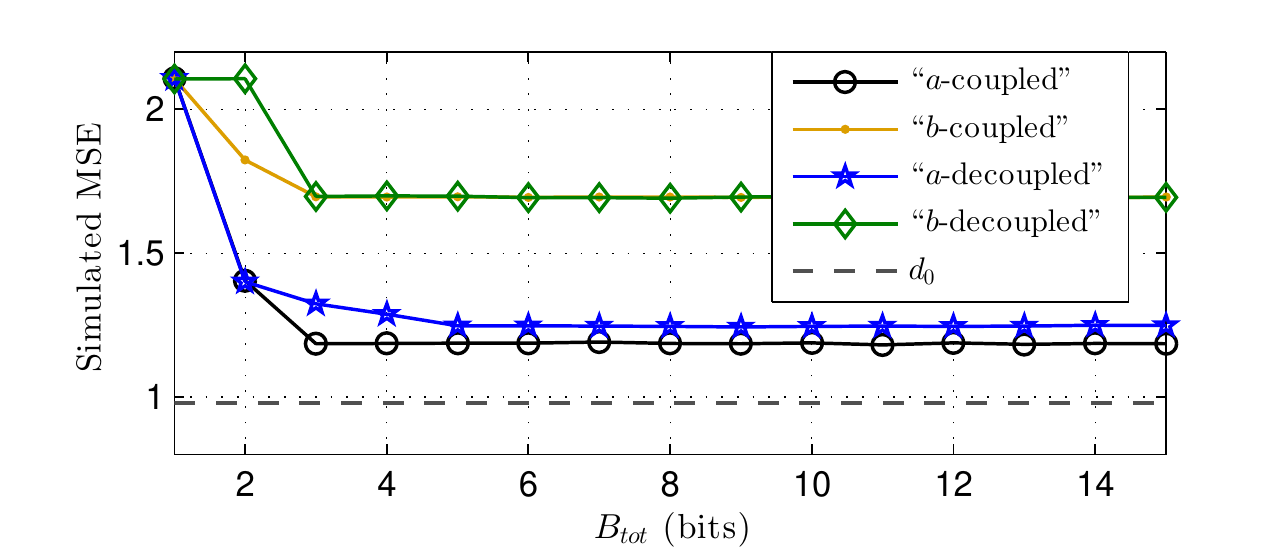}}
                                  \vspace{-.1cm}

        \caption{Simulated MSE vs. $B_{tot}$ for all algorithms}   
         \label{MSE-simulated-vs-B_tot} 
       
    \end{figure}
    
\section*{Acknowledgment}
This research is supported by NSF under grants CCF-1336123, CCF-1341966, and CCF-1319770.
\bibliographystyle{IEEEtran}
\bibliography{myref}

\end{document}